\documentclass[11pt,sort&compress]{article}
\usepackage{fullpage}
\usepackage{cite}
\usepackage{slashed}
\usepackage{graphicx}
\usepackage{amsmath}
\usepackage[english]{babel}
\usepackage{color}
\title{Top polarization studies in $H^-t$ and $Wt$ production}
\date{today}
\renewcommand{\vec}[1]{\mbox{\boldmath$ #1 $}}

\begin{document}
\bibliographystyle{utphys}
\newcommand{\msbar}{\ensuremath{\overline{\text{MS}}}}
\newcommand{\DIS}{\ensuremath{\text{DIS}}}
\newcommand{\abar}{\ensuremath{\bar{\alpha}_S}}
\newcommand{\bb}{\ensuremath{\bar{\beta}_0}}
\newcommand{\rc}{\ensuremath{r_{\text{cut}}}}
\newcommand{\Nd}{\ensuremath{N_{\text{d.o.f.}}}}
%\setlength{\parindent}{0pt}
%%%%%%%%%%%%%%%%%%%%%%%%%%%%%%%%%%%
\newcommand{\todo}[1]{\textbf{\textcolor{red}{(#1)}}}
%%%%%%%%%%%%%%%%%%%%%%%%%%%%%%%%%%%

\titlepage
\begin{flushright}
Nikhef-2012-020\\
LAPTH-059/12
\end{flushright}

\vspace*{0.5cm}

\begin{center}
{\Large \bf Top Polarization in Stop Production at the LHC}

\vspace*{1cm}
\textsc{G. B\'elanger$^a$\footnote{belanger@lapp.in2p3.fr}, R. M. Godbole$^{b}$\footnote{rohini@cts.iisc.ernet.in}, L. Hartgring$^{c}$\footnote{l.hartgring@nikhef.nl}, I. Niessen$^{d}$\footnote{i.niessen@science.ru.nl}} \\

\vspace*{0.5cm} $^a$ LAPTH, Univ. de Savoie, CNRS, B.P.110, F-74941\\
Annecy-le-Vieux Cedex, France

\vspace*{0.5cm} $^b$ Center for High Energy Physics, Indian Institute of 
Science\\ Bangalore 560 012, India\\

\vspace*{0.5cm} $^c$ Nikhef, Science Park 105, 1098 XG \\Amsterdam, The 
Netherlands\\

\vspace*{0.5cm} $^d$ Theoretical High Energy Physics, IMAPP, Faculty of Science, Mailbox 79,\\
P.O. Box 9010, NL-6500 GL Nijmegen, The Netherlands\\

 \end{center}

\vspace*{0.5cm}

\begin{abstract}

 We survey the expected  polarization of the top produced in the decay
 of a scalar top quark, $\tilde t \rightarrow {\tilde t} \, \,\chi_i^0,
 i =1-2$. The phenomenology is quite interesting, since the expected
 polarization depends both on the mixing in the stop and neutralino
 sectors and on the mass differences between the stop and the
 neutralino. We find that  a mixed stop  behaves almost  like a right-handed stop due to the larger hypercharge that enters the stop/top/gaugino coupling  and that these
polarisation effects disappear, when $m_{\tilde t_1} \approx m_t+m_{\tilde\chi^0_i}$.
 After
  a discussion on the expected top polarization from the decay of a
 scalar top quark, we focus on the interplay of polarization and
  kinematics at the LHC. We discuss different probes of the top
 polarization in terms of lab-frame observables. We find that these
  observables faithfully reflect the polarization of the parent
 top-quark, but also have a non-trivial dependence on the  kinematics
  of the stop production and decay process. In addition, we illustrate
 the effect of top polarization on the energy and transverse momentum
 of the decay lepton in the {\it laboratory} frame. Our results show
 that both spectra are softened substantially in case of a negatively
 polarized top, particularly for a large mass difference between the
 stop and the neutralino. Thus, the search strategies, and the
 conclusions that can be drawn from them, depends  not just on the
mass difference  $m_{\tilde t} - m_{\tilde  \chi_{i}^{0}}$ due to the usual
kinematic effects but also on the effects of top polarization on the
decay kinematics the extent  of which depends in turn on the said mass
difference.

\end{abstract}

\vspace*{0.5cm}

\section{Introduction}\label{s:intro}

The observation of a new boson at the LHC with properties broadly consistent with those of a Higgs
boson expected in the Standard  Model (SM),  by both the ATLAS and CMS collaborations~\cite{ATLAS:2012gk,CMS:2012gu},
indicates that the process of establishing the last missing piece of the SM has now begun. In spite of the great success of the SM, which would be crowned by this discovery, there are a number of observational issues that point at the need for physics beyond the SM (BSM). In particular, dark matter (DM) and baryon asymmetry in the Universe (BAU). In
addition, there are theoretical reasons for BSM physics, such as the instability of the EW scale under radiative corrections or a lack of fundamental
understanding of the observed wide range of the fermion masses. Supersymmetry (SUSY)~\cite{Wess:1974tw,Nilles:1983ge} has been one of the favourite
candidates for BSM physics, as it can provide a very elegant solution to many of these open questions, particularly
significant being the prediction of {\it at least one, low mass} Higgs boson, possibly the resonance that has been observed. 
Searches for light-flavoured squarks and gluinos at the Large Hadron Collider (LHC) have so far come up 
empty~\cite{Aad:2011ib,ATLAS-CONF-2012-109,Chatrchyan:2011zy,Chatrchyan:2012jx}. A key feature of almost all SUSY models is that
masses of all the supersymmetric particles depend crucially on the scale and the nature of the supersymmetry breaking mechanism, but the upper limit on the lightest Higgs boson mass depends only mildly on it. The only general theoretical pointers we have to the expected mass scales for SUSY breaking, and hence of the sparticle masses, come from naturalness arguments~\cite{Fischler:1981zk,Kaul:1981tp,Kaul:1981hi}. In SUSY, the low mass of the observed resonance, is naturally stable under large radiative corrections, provided the supersymmetry breaking scale is not too large. In particular,
the gluinos and most squarks can be quite heavy, as long as the top squark, or stop, is relatively light so that SUSY
has  a solution to offer to the  hierarchy problem as suggested originally~\cite{Kaul:1981wp,Dimopoulos:1981zb}. The upper limit
on the allowed stop masses for a given Higgs mass depends on the amount of fine tuning that is
tolerated \cite{Barbieri:1987fn,deCarlos:1993yy,Brust:2011tb}.

The recent Higgs results \cite{ATLAS:2012gk,CMS:2012gu}   
suggest, in the context of SUSY, a Higgs boson mass quite close to the upper bound on the mass of the lightest Higgs state. This points towards at least one relatively heavy stop~\cite{Arbey:2012dq,Barger:2012hr}, which naturally leads us to consider models with one light stop and at least one light neutralino, which is then the Lightest Supersymmetric Particle (LSP). This is the minimal 'light' SUSY particle content that one needs in order  to account for the observational hints of BSM physics such as DM and BAU. It is therefore particularly interesting to investigate  possibilities of such a light stop search at the LHC. 

Two points are to be noted. Due to the large mass of the top quark, the limits on squark masses obtained from the generic missing $E_{T} + jets (leptons)$ 
search~\cite{Aad:2011ib,ATLAS-CONF-2012-109, 
Chatrchyan:2011zy,Chatrchyan:2012jx}
are not directly applicable, even if one were to look at the limits
on the masses of light flavoured squarks produced 'directly'. Secondly, while it is true that the cross-section for the direct stop pair production is much smaller than the total squark-gluino cross-section, 
direct stop pair production processes are an interesting channel for stop searches, in view of the current constraints on the gluino mass. 
For example at $\sqrt{s}=8$~TeV the direct stop cross section at NLL level is  $\sim 85$ fb for $m_{\tilde t} = 500$ GeV,~\cite{Beenakker:1996ed,Beenakker:2010nq,Beenakker:2011fu,NLLfast} a value for the stop mass that is  currently allowed  by the data.

The third generation sfermion sector has always been a subject of great interest in sparticle phenomenology~\cite{sparticles}. In view of the above discussion, it is 
also clear why it has received even extra attention in both phenomenological ~\cite{Brust:2011tb,Desai:2011th,He:2011tp,Drees:2012dd,Berger:2012ec,Plehn:2012pr,Han:2012fw,
Berger:2012an,Bhattacherjee:2012ir} and experimental investigations.
Results on stop searches in direct stop pair production
have been presented both by the 
ATLAS~\cite{stop-atlas-prl-1:2012si,stop-atlas-prl-2:2012ar,Aad:2012uu,Aad:2012yr}, and the CMS \cite{CMS-PAS-SUS-12-023,
CMS-PAS-SUS-11-030,Chatrchyan:2012sv,Weber:1479435} collaborations. However, the interpretation of these searches has  some model-dependence and usually limits are quoted in simplified models. In any case, present data allows for top squarks well below the TeV scale. 

One new aspect of the stop search phenomenology is the possible presence of a top quark with possibly non-zero polarization in the resulting final state.  Since the top quark decays before it hadronizes, the polarization  can have implications for the kinematic distributions of the decay products and hence on the search strategies of the stop. If a stop is discovered, 
the top polarization can play a role in determining the 
properties of the stop and light neutralino.
In this paper, we investigate the longitudinal polarization of the top quark that results from stop decay;
\begin{align}
\tilde t_1\to t\,\tilde\chi_i^0\label{eq:stopdecay},
\end{align}
where $\tilde\chi_i^0, i =1,4$ stand for the four neutralinos. 
It has been shown~\cite{Nojiri:1994it} that the 
fermions produced in sfermion decays can have non-zero  polarization, which can depend on  the mixing in the  sfermion sector as well as the neutralino-chargino sector. It also depends on 
the velocity of the produced top quark and hence on the 
mass differences.  
 
Suggestions for using the polarization of heavy fermions as a probe of new physics models abound in literature, (see Ref.~\cite{Godbole:2011vw} and references therein for a recent summary). 
%we repeat this 2 paragraphs later (literally) and I think it fits better there. Top polarization is indeed a very useful observable as a probe of new physics, as it is sensitive to the helicity structure of the production process.  The bulk of top production at the LHC happens via the SM processes, which lead to unpolarized top quarks. The chiral nature of SUSY then indicates that top polarization can play a particularly useful role.
%This was supposed to be deleted as this was indeed copy pasted there if I recall right:-) RG
For example, in the $R$-parity violating MSSM, polarized top quarks can arise in the hadronic production of $t \bar t$ pair via a $t$--channel exchange of a
stau/stop~\cite{Hikasa:1999wy,Li:2006he} or in associated production of a slepton with a $t$ quark~\cite{Arai:2010ci}. Different BSM explanations of the top forward-backward asymmetry observed at the Tevatron, among them those involving $t$ channel exchange of a color singlet and a color octet scalar, can be discriminated using top 
polarization~\cite{Cao:2010nw,Jung:2010yn,Choudhury:2010cd,Krohn:2011tw}. Similarly, use of the top polarization to probe the mixing in the squark sector for the third generation squarks at $e^{+}e^{-}$ colliders has been a subject of a lot of detailed investigations\cite{Boos:2003vf,Gajdosik:2004ed}. At the $e^{+}e^{-}$ colliders the $\tilde t_{i} \tilde t_{j}^{*}$ production cross-sections also depend on the mixing in the stop sector. The joint measurements of the cross-sections and top polarization can then be used to reconstruct the parameters of the third generation squark sector. Of course at a collider like the LHC, in an $R$-parity conserving SUSY scenario, the production cross-sections do not depend on the mixing in the stop sector and  hence it is only the polarization which can provide a handle on it. 

Some aspects of top polarization in stop decay and observables for its measurements for the heavily boosted tops were discussed in 
~\cite{Shelton:2008nq}. Monte Carlo investigations of the 
top polarization expected in the decay of a light  stop quark 
($\sim 300$--$500$ GeV) following direct stop pair production for $14$ TeV LHC, along with its possible measurements in the effective top rest frame with a view to extract an effective top mixing angle, have been  carried out in~\cite{Perelstein:2008zt}. More recently, an observable for top polarization in terms of the energy fraction of decay leptons, in events containing $t \bar t$ pair and missing $E_{T}$ was studied for a light stop $\sim 300$ -- $400$ GeV, at the $8$ TeV LHC~\cite{Berger:2012an}. 
Ref.~\cite{Bhattacherjee:2012ir} has explored the possibility 
of getting information on the  top polarization and hence on the stop mixing angle at the 14 TeV LHC, including detector level effects, using the hadronic decay of the boosted top and jet substructure methods for measurement of the top polarization~\cite{Krohn:2009wm}. Experimental explorations of the top polarization  at the LHC in $t \bar t$ events, using the angular distributions  of the decay products of the top in the reconstructed top rest frame have now begun~\cite{ATLASCONF-133}.

Top polarization is indeed a very useful observable as a probe of new physics at the LHC as  it
is sensitive to the helicity structure of the production process and  the bulk of top production at the LHC happens  via the SM processes which lead to unpolarized top quarks. 
Due to the large mass of the top quark, its polarization is also amenable to  experimental determination quite well through  a study of its leptonic decay products. There is a strong correlation between the polarization of the top quark and the angular distributions of its decay leptons. This correlation is not affected by higher-order corrections \cite{Jezabek:1988ja,Czarnecki:1990pe,Brandenburg:2002xr} or new physics contributions \cite{Grzadkowski:1999iq,Grzadkowski:2002gt,Grzadkowski:2001tq,Hioki:2002vg,Ohkuma:2002iv,Rindani:2000jg,Godbole:2002qu} to the decay. Angular distributions of the decay leptons provide therefore a robust probe of the top polarization and hence of the new physics.

The aim of  this paper is to present in detail the 
dependence of the expected top polarization from stop decay on the 
mixings in the stop and chargino/neutralino sectors, as well as on the mass differences between the stop and neutralinos. We will present results in terms of the relevant supersymmetric parameters that are still allowed in view of the LHC results. This has a two-fold purpose. Firstly, it gives us a pointer to the possible kinematic effects that 
this top polarization can have on its decay products and 
hence to the implications of this feature for  the search strategies for the stop which use final  states containing a top quark.   The second is to explore how measurement of the longitudinal 
polarization of  the resulting top quark can  be used to help determine the properties of the stop and the light neutralinos, after  the discovery. To that end,  we study observables 
of the top polarization, at the $8$ TeV LHC in terms of 
the kinematic variables of the decay lepton in the laboratory 
frame that have been suggested 
earlier~\cite{Godbole:2006tq,Godbole:2010kr,Godbole:2009dp,Godbole:2011vw}.

In this paper, we will first discuss in 
section~\ref{s:polarization} how the polarization of the top 
is affected by the properties of the stop and the neutralinos. 
We then study possible top polarizations by scanning the 
relevant SUSY parameters in section~\ref{s:scan}. In 
section~\ref{s:observables} we examine polarization-dependent kinematic variables in the laboratory frame for specific 
benchmark points and discuss possible observables for the polarization constructed out of the angular variables. We 
conclude in section~\ref{s:conclusion}.

\section{Top Polarization from Stop Decay}\label{s:polarization}

We begin by briefly recalling the correlation between the top quark spin and the flight direction of the charged lepton in the decay. 
When determining the polarization of the top, we consider top quark decays that produce a charged lepton $l^+$, which we take to be an electron or a muon
\begin{align}
t\rightarrow W^+b\rightarrow l^+ \nu_l\, b\,.\label{eq:topdecay}
\end{align}
For simplicity here and in what follows, we ignore off-diagonal elements in the CKM matrix and we only consider top quarks, which can be distinguished from anti-top quarks using the charge of the lepton. As mentioned in the introduction, the top polarization is sensitive to the production process, not to corrections to its decay. To see this~\cite{Godbole:2006tq}, let us employ the Narrow Width Approximation (NWA) for the top quark. This allows us to split the spin-averaged matrix element squared $\overline{|{\cal M}|^2}$ into a part $\rho(\lambda,\lambda')$ that corresponds to the production  of the top quark, and a part $\Gamma(\lambda,\lambda')$ that corresponds to its decay
\begin{align}
\overline{|{\cal M}|^2}=\frac{\pi\delta(p_t^2-m_t^2)}{\Gamma_tm_t}\sum_{\lambda,\lambda'}\rho(\lambda,\lambda')\Gamma(\lambda,\lambda')\,.
\end{align}
Here $p_t^\mu$, $m_t$ and $\Gamma_t$ are the top quark momentum, mass and total decay width respectively, while $\rho(\lambda,\lambda')$ and $\Gamma(\lambda,\lambda')$ are matrices given by
\[\rho(\lambda,\lambda')={\cal M}_\rho(\lambda){\cal M}^*_\rho(\lambda')\quad\mbox{and}\quad \Gamma(\lambda,\lambda')={\cal M}_\Gamma(\lambda){\cal M}^*_\Gamma(\lambda')\,,\]
with ${\cal M}_\rho(\lambda)$ the matrix element of the production of a  top quark with helicity $\lambda$ and ${\cal M}_\Gamma(\lambda)$ the corresponding decay amplitude. To obtain the averaged matrix element squared $\overline{|{\cal M}|^2}$, we have to sum over the helicities $\lambda$ and $\lambda'$. However, we can also project on these helicities to obtain the polarized cross section. To this end, we define top polarization vectors $S^a$ that form, together with the top momentum, an orthogonal set and are normalized to $S^a\cdot S^b=-\delta^{ab}$. We can then perform the helicity projection using the identities \cite{Bouchiat1958416,springerlink:10.1007/BF03026451}:
\begin{align}
u(p_t,\lambda')\bar u(p_t,\lambda)=\frac{1}{2}\big(\delta_{\lambda\lambda'}+\gamma_5\slashed S^a\tau^a_{\lambda\lambda'}\big)(\slashed p_t+m_t)\,,\\
v(p_t,\lambda')\bar v(p_t,\lambda)=\frac{1}{2}\big(\delta_{\lambda\lambda'}+\gamma_5\slashed S^a\tau^a_{\lambda\lambda'}\big)(\slashed p_t-m_t)\,,
\end{align}
with $\tau^a$ the Pauli matrices. Since the transverse polarization is generally small, we will only consider the longitudinal polarization vector $S^3$. Its spatial part is chosen to be parallel to the top three-momentum, leading to
\begin{align}
S^3=\frac{1}{m_t}\big(|\vec p_t|,E_t\hat{\vec p}_t\big)\,.\label{eq:spinvector}
\end{align}
Note that $S^3$ is not a Lorentz vector, reflecting the fact that the top quark helicity is not a Lorentz-invariant quantity. The top polarization is then defined as
\begin{align}
P_t=\frac{\sigma(+,+)-\sigma(-,-)}{\sigma(+,+)+\sigma(-,-)}\,,
\label{eq:Ptdef}
\end{align}
where $\sigma(+,+) \left(\sigma(-,-)\right)$ is the cross section for a positive (negative) helicity top quark. A negative (positive) polarization therefore corresponds to a left-handed (right-handed) top quark. In~\cite{Boos:2003vf} it was shown for a top quark originating from the decay~\eqref{eq:stopdecay}, the following expression for the polarization holds 
\begin{align}
P_t(\tilde t_1\to t\,\tilde\chi_i^0)=\frac{\big((G^R_i)^2-(G^L_i)^2\big)f_1}{(G^R_i)^2+(G^L_i)^2-2G^R_iG^L_if_2}\,,\label{eq:topstoppol}
\end{align}
where $f_1$ and $f_2$ are kinematical factors which in the stop rest frame reduce to
\begin{align}
f_1&=\frac{m_t(p_{\tilde\chi}\!\cdot\! S^3)}{(p_t\!\cdot\! p_{\tilde\chi})}\to\frac{\lambda^{\frac{1}{2}}(m_{\tilde t}^2,m_t^2,m_{\tilde\chi}^2)}{m_{\tilde t}^2-m_t^2-m_{\tilde\chi}^2}\,,&
f_2&=\frac{m_tm_{\tilde\chi}}{(p_t\!\cdot\! p_{\tilde\chi})}\to\frac{2m_tm_{\tilde \chi}}{m_{\tilde t}^2-m_t^2-m_{\tilde\chi}^2}\,,\label{eq:fs}
\end{align}
with $\lambda(x,y,z)=x^2+y^2+z^2-2xy-2xz-2yz$ the K\"all\'en function. The quantities $G^L_i$ and $G^R_i$ are the stop couplings to the neutralino $\tilde\chi_i^0$ and a left- or right-handed top respectively. If we ignore again mixing in the flavour sector and choose the mixing matrices to be real, they are given by \cite{sparticles}
\begin{eqnarray}
G^L_i&=&-\sqrt{2}g_2\bigg(\frac{1}{2}Z_{i2}+\frac{1}{6}\tan\theta_WZ_{i1}\bigg)\cos\theta_{\tilde t}-\frac{g_2m_t}{\sqrt{2}M_W\sin\beta}Z_{i4}\sin\theta_{\tilde t}\,,\label{eq:stopcouplings}\\\nonumber
G^R_i&=&\frac{2\sqrt{2}}{3}g_2\tan\theta_WZ_{i1}\sin\theta_{\tilde t}-\frac{g_2m_t}{\sqrt{2}M_W\sin\beta}Z_{i4}\cos\theta_{\tilde t}\,,
\label{eq:couplings}
\end{eqnarray}
where $g_2$ is the SU(2)$_L$ gauge coupling, $\theta_W$ is the weak mixing angle and $M_W$ is the $W$ mass. The polarization then depends on the SUSY parameters through the neutralino mixing matrix $Z$, the stop mixing angle $\theta_{\tilde t}$ and the ratio of the two Higgs vacuum expectation values, $\tan\beta$. Moreover it is clear from Eq.~\eqref{eq:topstoppol} that the top polarization is affected by the masses involved and perhaps less obviously by the stop boost. Let us now discuss these effects in turn.

\subsection{Stop and Neutralino Mixing}

The top polarization Eq.~\ref{eq:Ptdef} depends on the couplings $G^{L,R}_i$, Eq.~\eqref{eq:stopcouplings}, which contain the stop mixing $\theta_{\tilde t}$ and neutralino mixing. The mixing $\theta_{\tilde t}$ results from the diagonalization of the stop mass matrix in the $L-R$ basis, leading to the mass eigenstates $\tilde t_1$ and $\tilde t_2$
\begin{align}
M_{\tilde t}^2=
\left(\!\begin{array}{cc}
m_{\tilde t_L}^2+\Delta_L+m_t^2&-m_t(A_t+\mu\cot\beta)\\
-m_t(A_t+\mu\cot\beta)&m_{\tilde t_R}^2+\Delta_R+m_t^2
\end{array}\!\right),\quad
\left(\!\begin{array}{c}
\tilde t_1\\\tilde t_2
\!\end{array}\right)=
\left(\!\begin{array}{cc}
\cos\theta_{\tilde t}&\sin\theta_{\tilde t}\\
-\sin\theta_{\tilde t}&\cos\theta_{\tilde t}
\end{array}\!\right)
\left(\!\begin{array}{c}\tilde t_L\\\tilde t_R\end{array}\!\right),
\end{align}
with $m_{\tilde t_{L,R}}$ the soft masses of the left- and right-handed stop, $A_t$ the top trilinear coupling, $\mu$ the Higgs mass parameter, and $\Delta_L=(\tfrac{1}{2}-\tfrac{2}{3}\sin\theta_W^2)M_Z^2\cos2\beta$, $\Delta_R=(\tfrac{2}{3}\sin\theta_W^2)M_Z^2\cos2\beta$ with $M_Z$ the $Z^0$ mass.

The neutralino mixing matrix, $Z$ is determined by the diagonalization of the neutralino mass matrix $M_n$: 
\begin{align}
M_n=
\left(\!\begin{array}{cccc}
M_1&0&-M_Zc_\beta s_W&M_Zs_\beta s_W\\
0&M_2&M_Zc_\beta s_W&-M_Zs_\beta c_W\\
-M_Zc_\beta s_W&M_Zc_\beta c_W&0&-\mu\\
M_Zs_\beta s_W&-M_Zs_\beta c_W&-\mu&0
\end{array}\!\right),\quad
\left(\!\begin{array}{c}\tilde \chi_1^0\\\tilde \chi_2^0\\\tilde \chi_3^0\\\tilde \chi_4^0\!\end{array}\right)=Z
\left(\!\begin{array}{c}\tilde B^0\\\tilde W^0\\\tilde h_1^0\\\tilde h_2^0\end{array}\!\right),
\end{align}
with $M_1$ and $M_2$ the bino and Wino gaugino masses, $s_W=\sin\theta_W$, $c_W=\cos\theta_W$, $s_\beta=\sin\beta$ and $c_\beta=\cos\beta$. Our subsequent investigations of the to ppolarization will be guided by a few salient aspects in this mixing, which we now discuss. 

Firstly, one notes that the strength of the bino($\tilde B$)  coupling to stop-top is proportional to the top hypercharge. As a result, a bino-like neutralino couples more strongly to the right-handed (RH) components than to the left-handed (LH) ones, yielding a more positive top polarization than one might naively expect from a given stop mixing.

Secondly, recall that the Wino $\tilde W$ only couples to the left-handed stop components, producing left-handed top quarks only. According to Eq.~\ref{eq:topstoppol}, a pure Wino thus always leads to $P_t=-f_1$ in the stop rest frame. As a result, polarization cannot be used to distinguish between different stop mixing for Wino-type neutralinos. In the rest of the paper we will thus limit ourselves to neutralinos with a small Wino component.

Thirdly, for the intermediate to large values of $\tan\beta$ that are allowed for the Higgs mass constraint, $\sin\beta\approx 1$, therefore the couplings in Eq.~\ref{eq:couplings}  hence the top polarization only mildly depend on $\tan\beta$.

\begin{figure}[!h]
\centering
\includegraphics[width=0.58\linewidth]{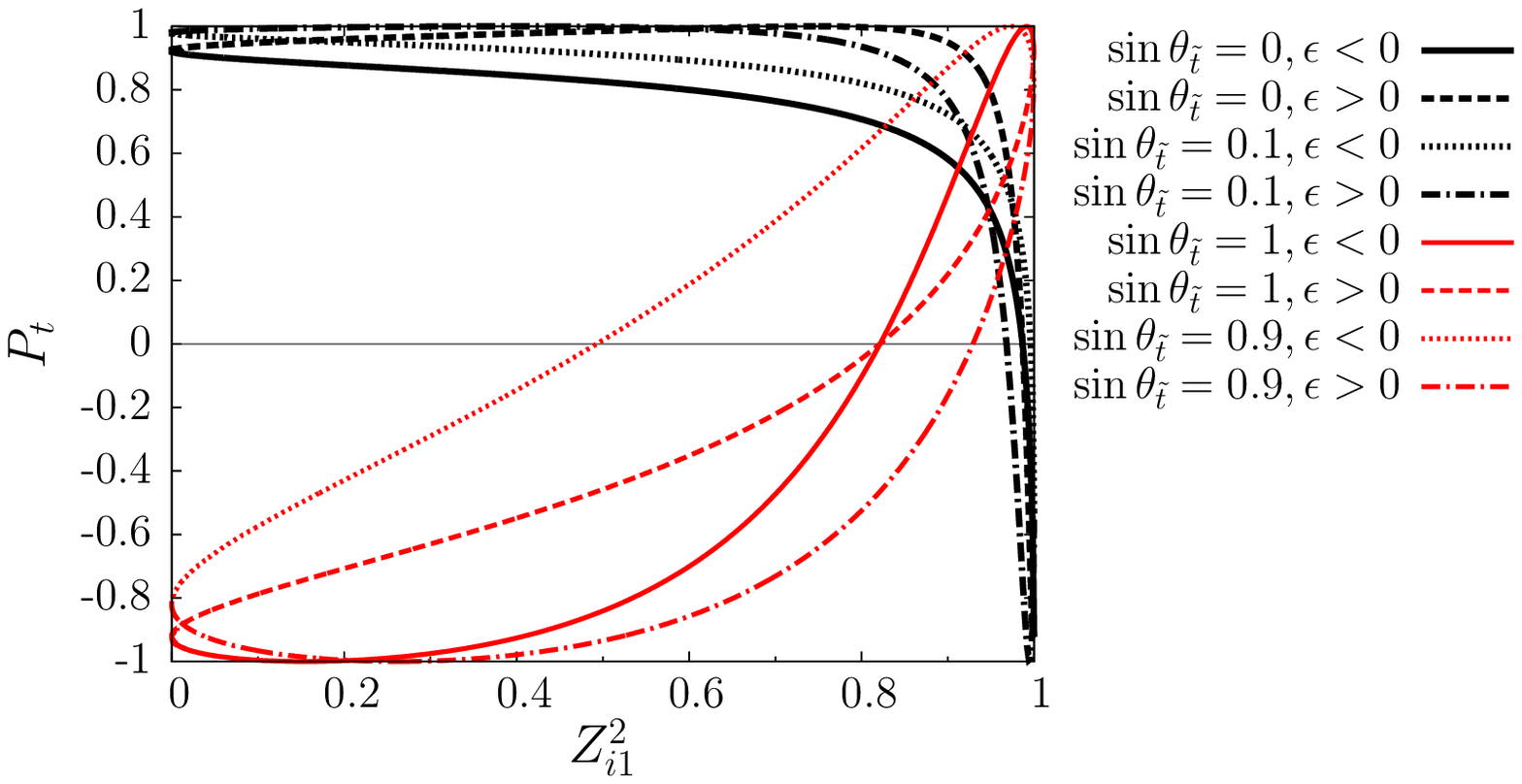}
\hfill
\includegraphics[width=0.4\linewidth]{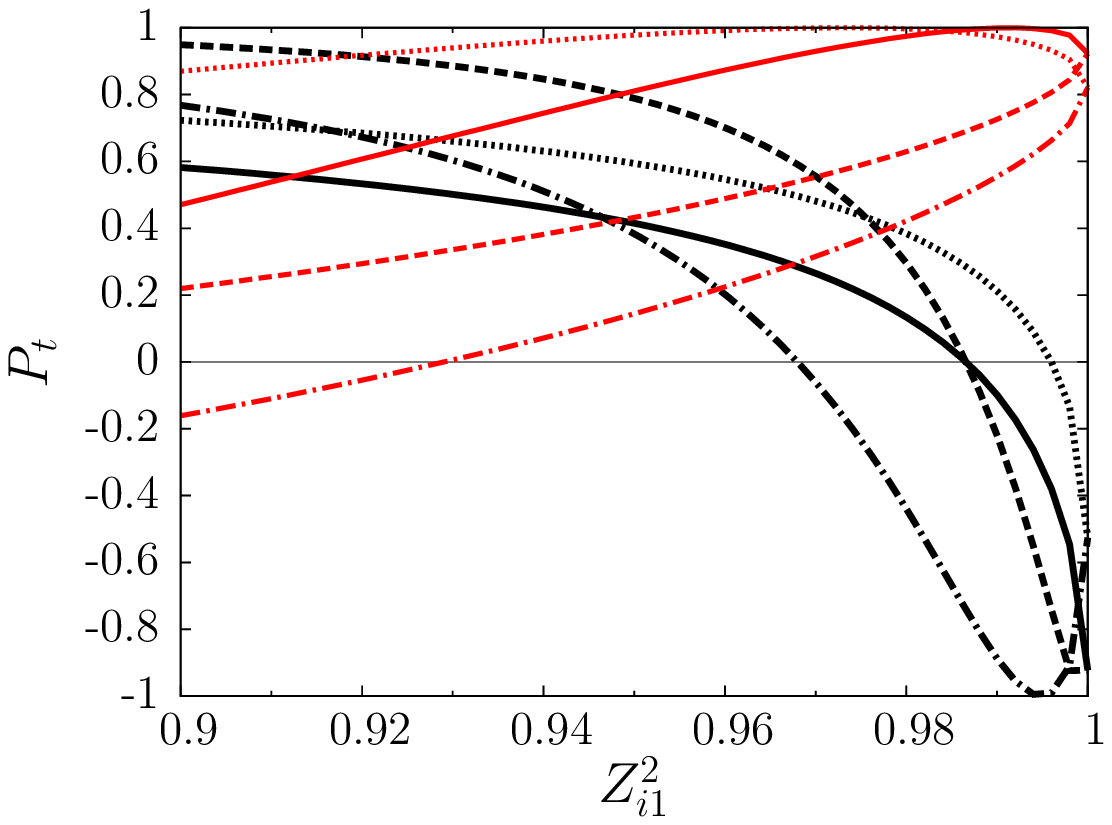}
\caption{Dependence of the top polarization on the neutralino content in the stop rest frame. The red thin lines correspond to right-handed stops, while the black thick lines correspond to left-handed stops. Results are shown for pure as well as slightly mixed stops, and for different signs of $\mu$. We have taken $Z_{i4}=\epsilon \sqrt{(1-Z_{i1}^2)}$, $\epsilon=\pm 1$ to approximate the higgsino-content for a given bino-content and have taken $m_t=173.1$ GeV, $m_{\tilde t}=500$ GeV, $m_{\tilde\chi}=200$ GeV and $\tan\beta=10$. The plot on the right shows the behaviour for high bino-content.}\label{fig:analytic_neutmix}
\end{figure}

Finally, the stop-top-neutralino coupling does not involve the first higgsino component $\tilde h_1^0$. Ignoring the Wino component, the key variables in the neutralino mixing matrix are thus the bino component $Z_{i1}$ and the second higgsino component $Z_{i4}$. The relative sign between the bino and the higgsino components can impact the polarization because of the term 
%depends on the sign of $\mu$  It is positive for $\mu<0$, while it is negative for $\mu>0$ \todo{is this always true?}. 
proportional to $G^R_iG^L_i$ in Eq.~\eqref{eq:topstoppol}. This can be seen in Fig.~\ref{fig:analytic_neutmix}, where the top polarization in the stop rest frame is plotted as a function of the bino content for both left- and right-handed stops. The figure on the right zooms into the region with high bino-content. The results are shown for both relative signs of $Z_{i1}$ and $Z_{i4}$ and also for stops that are not entirely left- or right-handed. 

The figure shows that in general the polarization behavior is as expected:  dominantly right-handed stops produce a negative top polarization when they decay to a higgsino, and a positive polarization when they decay to a bino. Left-handed stops have the opposite behaviour. Notice that is in correspondence to the first aspect mentioned above, for right-handed stops in particular, even a slight change in the stop mixing angle has a large effect on the polarization. We observe that the polarization for left-handed stops
is not very sensitive to the exact neutralino content when it is higgsino-like and that the polarization varies very rapidly from 1 to -1 for an almost pure bino. 
Moreover, the maximum polarization $P_t=\pm1$ cannot occur for a  decay into a pure bino or higgsino due to the mass effects in Eq.~\eqref{eq:topstoppol}. 
This effect becomes more pronounced for smaller stop-neutralino mass differences.

\begin{figure}[!h]
\centering
\includegraphics[width=0.58\linewidth]{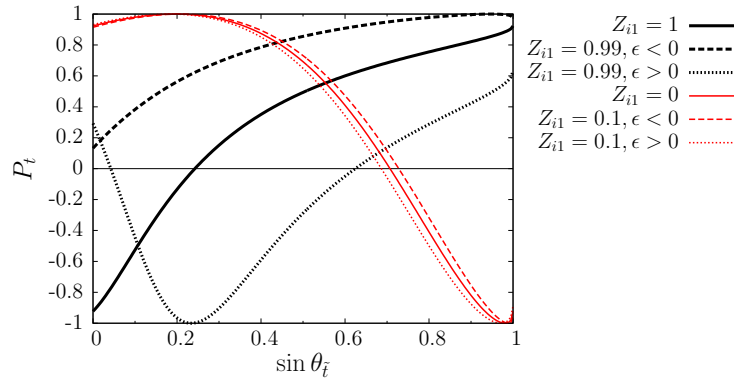}
\caption{Dependence of the top polarization on the stop mixing in the stop rest frame. The red thin lines correspond to higgsino-like neutralinos, while the black thick lines correspond to bino-like neutralinos. Results are shown for pure as well as slightly mixed neutralinos, and for different signs of $\mu$. We fix the parameters as in  Fig.~\ref{fig:analytic_neutmix}
}\label{fig:analytic_stopmix}
\end{figure}

For a complementary perspective we show in Fig.~\ref{fig:analytic_stopmix} the dependence of the top polarization on the stop mixing for a top quark that originates from a stop that is at rest.
For both the pure bino state and the dominantly  higgsino state, the polarization indeed behaves as one would expect from Eq.~\eqref{eq:topstoppol}.
As in Fig.~\ref{fig:analytic_neutmix}, we see that the polarization is very sensitive to small fluctuations in the bino component for $Z_{i1}\approx1$.   In this case, both terms in the $G_i^R$ coupling in Eq.~\ref{eq:couplings} become relevant, 
the first is suppressed by the stop mixing and the second by the higgsino mixing, hence the large fluctuation in the polarization 
for small values of $\sin\theta_{\tilde t}$. 

\subsection{Masses}\label{sec:Masses}

\begin{figure}[!h]
\centering
\includegraphics[width=0.4\linewidth]{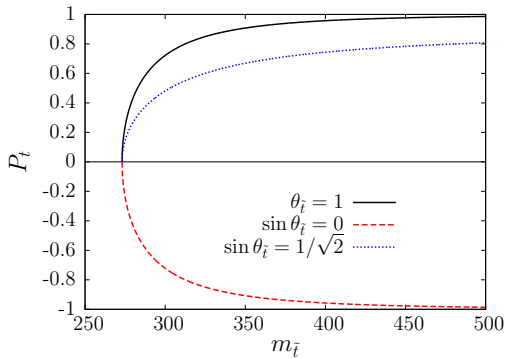}
\caption{Dependence of the top polarization in the stop rest frame on the stop-neutralino mass difference for a neutralino that is purely bino and different stop mixing. We have taken $m_t=173.1$ GeV, $m_{\tilde\chi}=100$ GeV and $\tan\beta=10$.}\label{fig:analytic_massdiff}
\end{figure}

We have already seen that the stop and neutralino masses influence the polarization. This effect is shown in Figure~\ref{fig:analytic_massdiff}.

We see that a small mass difference between the stop and the neutralino leads to a smaller polarization due to the $f_1$ and $f_2$ functions in Eq.~\eqref{eq:fs}. For mass differences of 200-300 GeV, this dependence is negligible. Note that the top originating from a completely mixed stop resembles a right-handed stop because of the effect of the hypercharge mentioned in the previous section.

Figure~\ref{fig:analytic_massdiff} only shows the results for the pure bino case, where the function $f_2$ does not contribute to the stop polarization~\eqref{eq:topstoppol}. We have seen in Figures~\ref{fig:analytic_neutmix} and \ref{fig:analytic_stopmix} that masses can have more intricate effects for mixed states due to the contribution of the $f_2$ function.

\subsection{Stop Boost}\label{sec:Stop_Boost}

\begin{figure}[!h]
\centering
\includegraphics[width=0.45\linewidth]{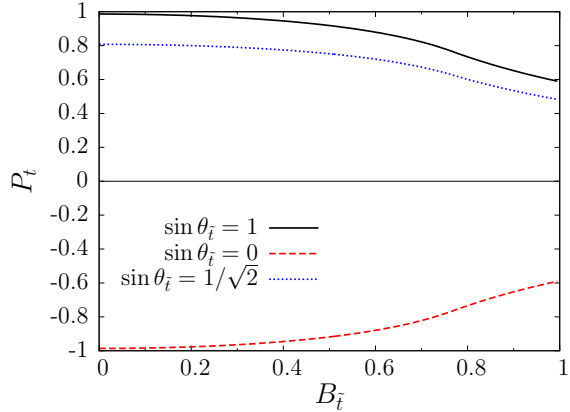}
\caption{The dependence of the top polarization on the stop boost for a neutralino that is purely bino and different stop mixing is shown. We have taken $m_t=173.1$ GeV, $m_{\tilde t}=500$ GeV, $m_{\tilde\chi}=100$ GeV and $\tan\beta=10$. }\label{fig:analytic_boostplot}
\end{figure}

So far we have studied the top polarization in the stop rest frame. However, as we can see from Eq.~\eqref{eq:spinvector}, the polarization vector $S^3$ is not a Lorentz vector. Thus the polarization is frame-dependent. We can quantify this effect using the stop boost
\begin{align}
B_{\tilde t}=\frac{|\vec{p}_{\tilde t}|}{E_{\tilde t}}\,.
\end{align}
The result is plotted in Figure~\ref{fig:analytic_boostplot}, showing that the polarization is reduced with increasing stop boost. The precise magnitude of the effect depends on the masses involved, since the boost-dependence of the result originates from the mass of the top quark. For the interesting stop and neutralino masses, the boost that the top obtains from the stop decay has the same order of magnitude as the stop boost. On average, these boosts tend to cancel each other, yielding a lower polarization.

\begin{figure}[!h]
\centering
\includegraphics[width=0.45\linewidth]{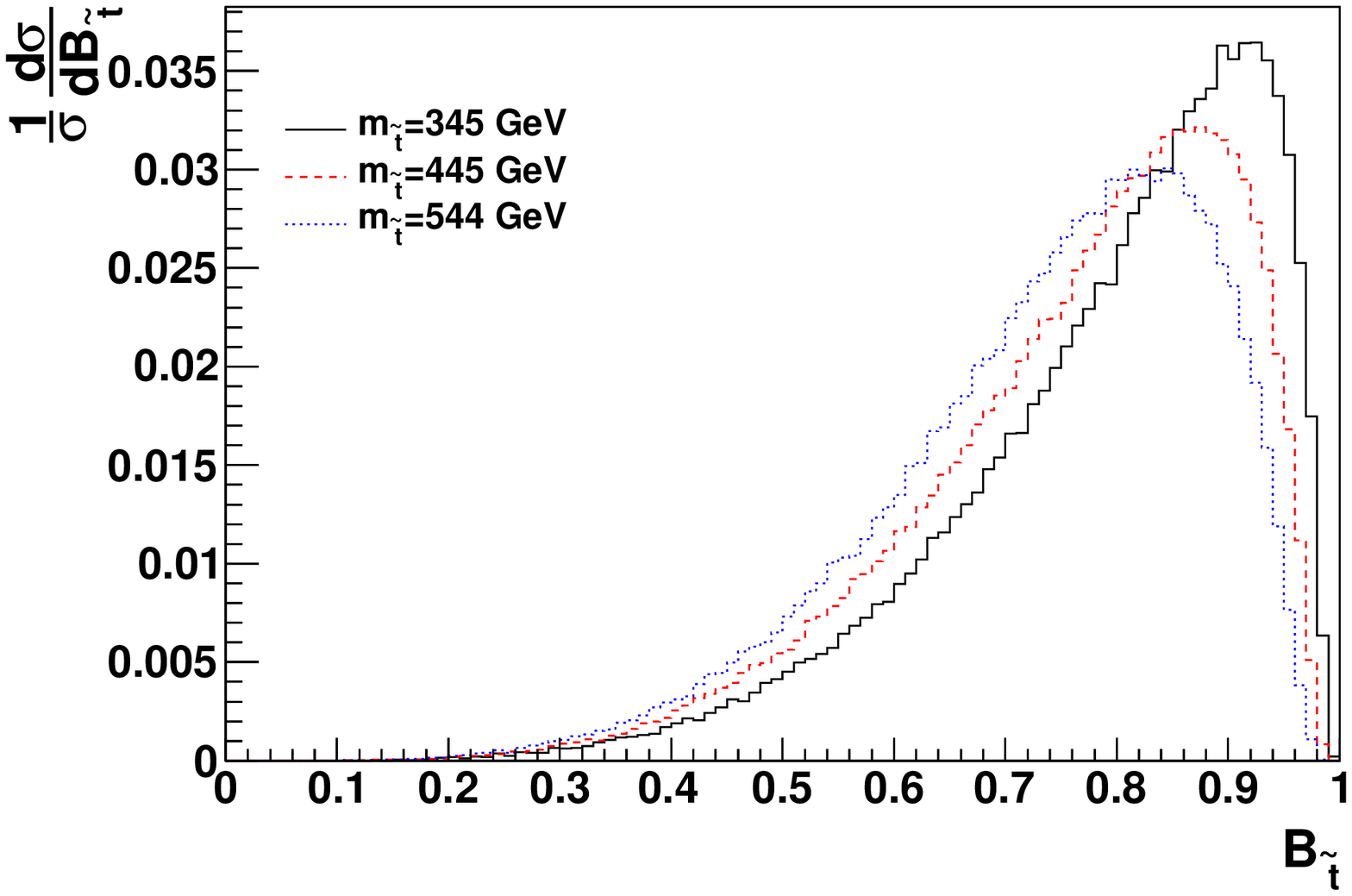}
\includegraphics[width=0.45\linewidth]{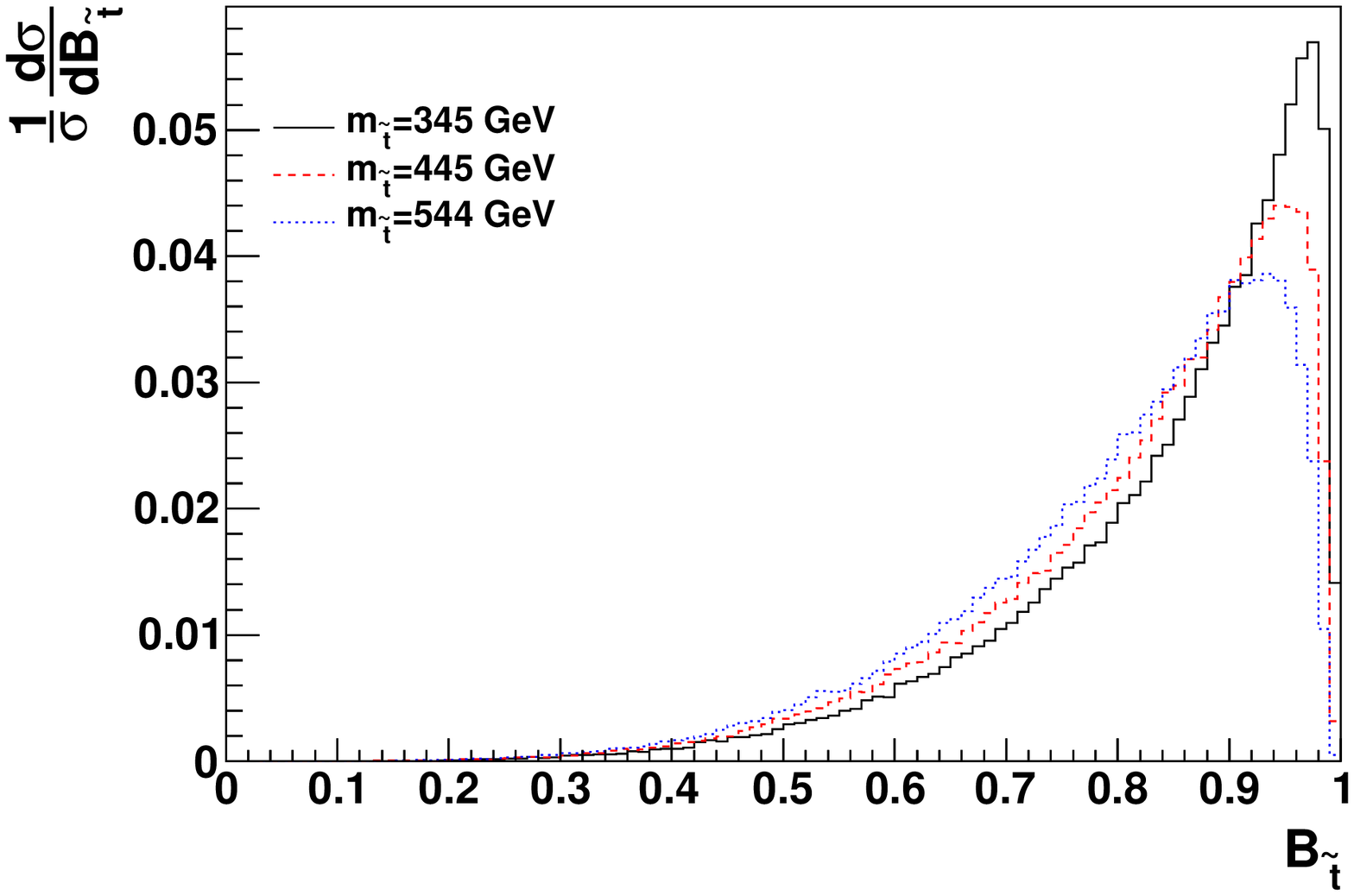}
\caption{The distribution of the stop boost at the LHC with an 8 TeV CM energy for different stop masses is shown on the left-hand side and 14 TeV CM energy on the right-hand side. Both distributions have been generated with Madgraph~\cite{Alwall:2011uj,Alwall:2007st}.}\label{fig:analytic_boostplot2}
\end{figure}

Although this sounds like a serious complication for studying the polarization at the LHC, the situation is in fact not that bad. The plots  of Figure~\ref{fig:analytic_boostplot2} shows the distribution of the stop boost at the LHC with a CM energy of 8 and 14 TeV. We see that within the relevant range of stop masses, the boost is fairly constant. Thus, the effect of the boost will reduce the polarization for all stop masses, but the explicit mass dependence due to the boost is small.

\section{Sensitivity to SUSY Parameters}\label{s:scan}

The top polarization in the stop rest frame is sensitive to the stop and neutralino masses and mixing. In the previous section, we have varied one of the relevant parameters at a time. In this section, we examine the dependence of the polarization on the MSSM parameters. We choose parameters such that the value of the light stop mass
is around 500~GeV.
This mass  leads to a large production cross section and has sufficient phase space for the stop to decay  in a top and a neutralino for a wide range of values for the neutralino mass. Furthermore, this mass satisfies the
 limits from direct stop production at the LHC 7 TeV.  For example the ATLAS Collaboration has excluded a stop up to nearly 500 GeV when the neutralino is massless, but provide no limit if the LSP is heavier than 150 GeV~\cite{stop-atlas-prl-2:2012ar}. 

We first choose fixed values for the soft parameters in the stop sector
and vary $M_1$ and $\mu$ to show the dependence on the neutralino composition. 
 The four sets of parameters  are given in Tab.~\ref{stop_sec_par}. We vary $M_1$ and $\mu$ to show the dependence on the neutralino composition. 
We   set $M_2=4M_1$ to decouple the wino-state and fix $M_3=1.5$~TeV, $M_A=1$~TeV. For the soft parameters in the sfermion sector, we choose
a common mass for  all sleptons $M_{\tilde l}=800$~GeV and for the first and second generation of squarks, $M_{\tilde q_i}=2$~TeV.  All trilinear couplings  except $A_t$ are set to zero. The supersymmetric spectrum and the Higgs masses are computed with SuSPect~\cite{Djouadi:2002ze}, which includes radiative corrections. 

At this point, we do not impose any constraints on the model. However, we choose the parameters of the stop sector such that the Higgs mass is within the measured range ($m_H=125.7\pm 0.4$~GeV, the average of CMS and ATLAS results~\cite{CMS:2012gu,ATLAS:2012gk})  for a large fraction of the parameter space explored after allowing for an additional 2-3 GeV theoretical uncertainty.  
The expectations for different observables from the flavour or dark matter sector are not taken into account at this point. They will be briefly discussed at the end of this section. 

\begin{table}[!h]
\centering
\begin{tabular}{l|cccc|cccc}
  & $M_{\tilde Q_3}$ (TeV) & $M_{\tilde u_3}$ (TeV) & $A_t$ (TeV) &$\tan\beta$&$M_{\tilde t_1}$ (GeV) & $\sin \theta_{\tilde t}$& $\cos \theta_{\tilde t}$&$M_h$(GeV)\\ \hline
 LH & $0.49$ & $2.00$ & $3.00$&10& 521. & -0.126&0.992  &126.4 \\
 XLH & $0.55$ & $1.40$ & $2.40$&20&510. & -0.223&0.975  &124.8 \\
 XRH & $1.05$ & $0.60$ & $1.88$&20& 498.& 0.946&-0.323  &124.0 \\
 RH & $2.00$ & $0.45$ & $2.40$ &10& 508. & 0.996&-0.095  &125.5 \\
\end{tabular}
\caption{Choices of parameters in the stop sector for two  mostly LH and two mostly RH stops. In each case we also consider a partly mixed light stop (XLH and XRH). The last columns specify  the  light stop mass, the stop mixing  and the Higgs mass  for
$|\mu|=  300 {\rm GeV}$, $M_1= 250 {\rm GeV}$.}
\label{stop_sec_par}
\end{table}

\begin{figure}[!h]
\centering
\includegraphics[width=0.49\linewidth]{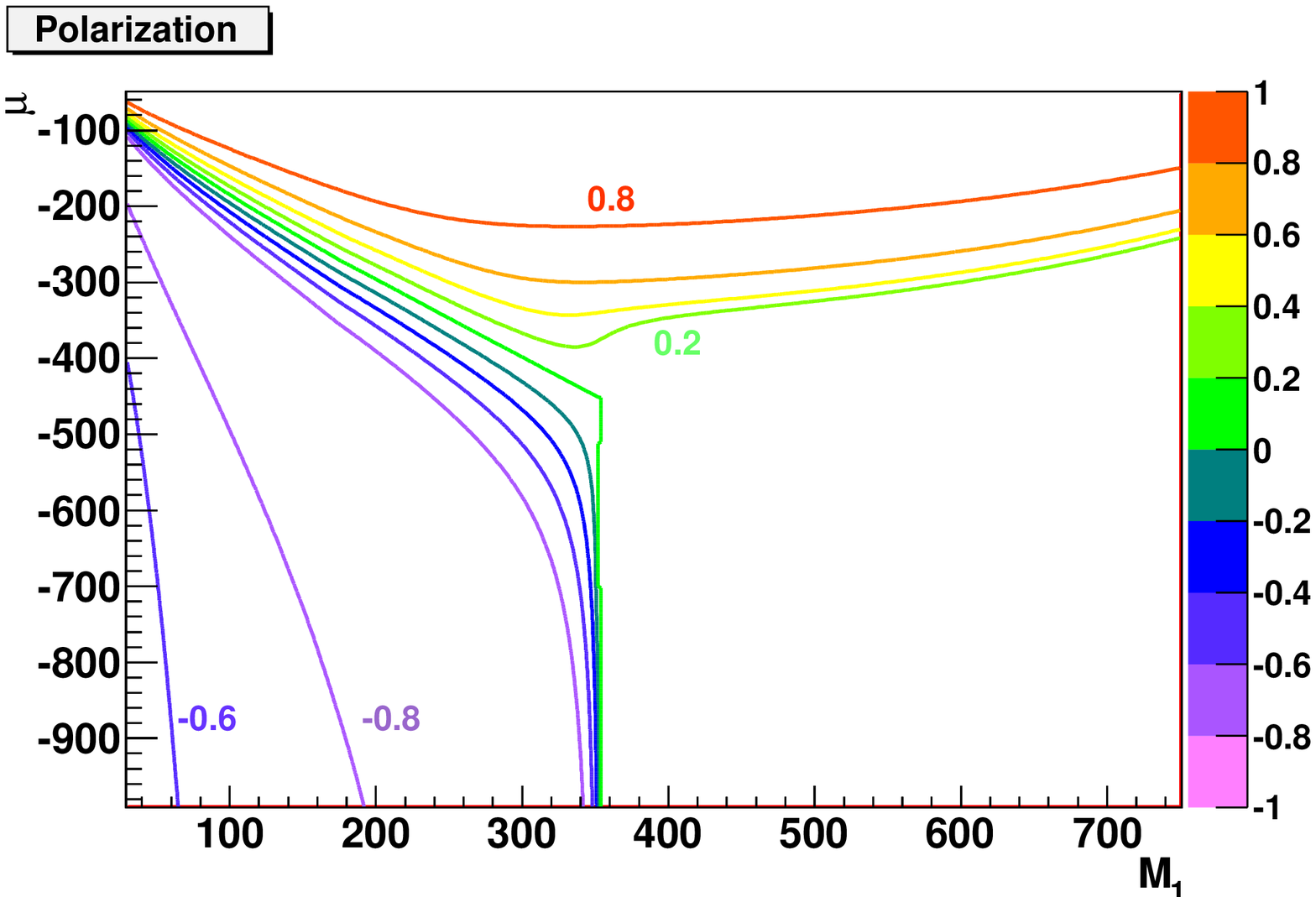}
\includegraphics[width=0.49\linewidth]{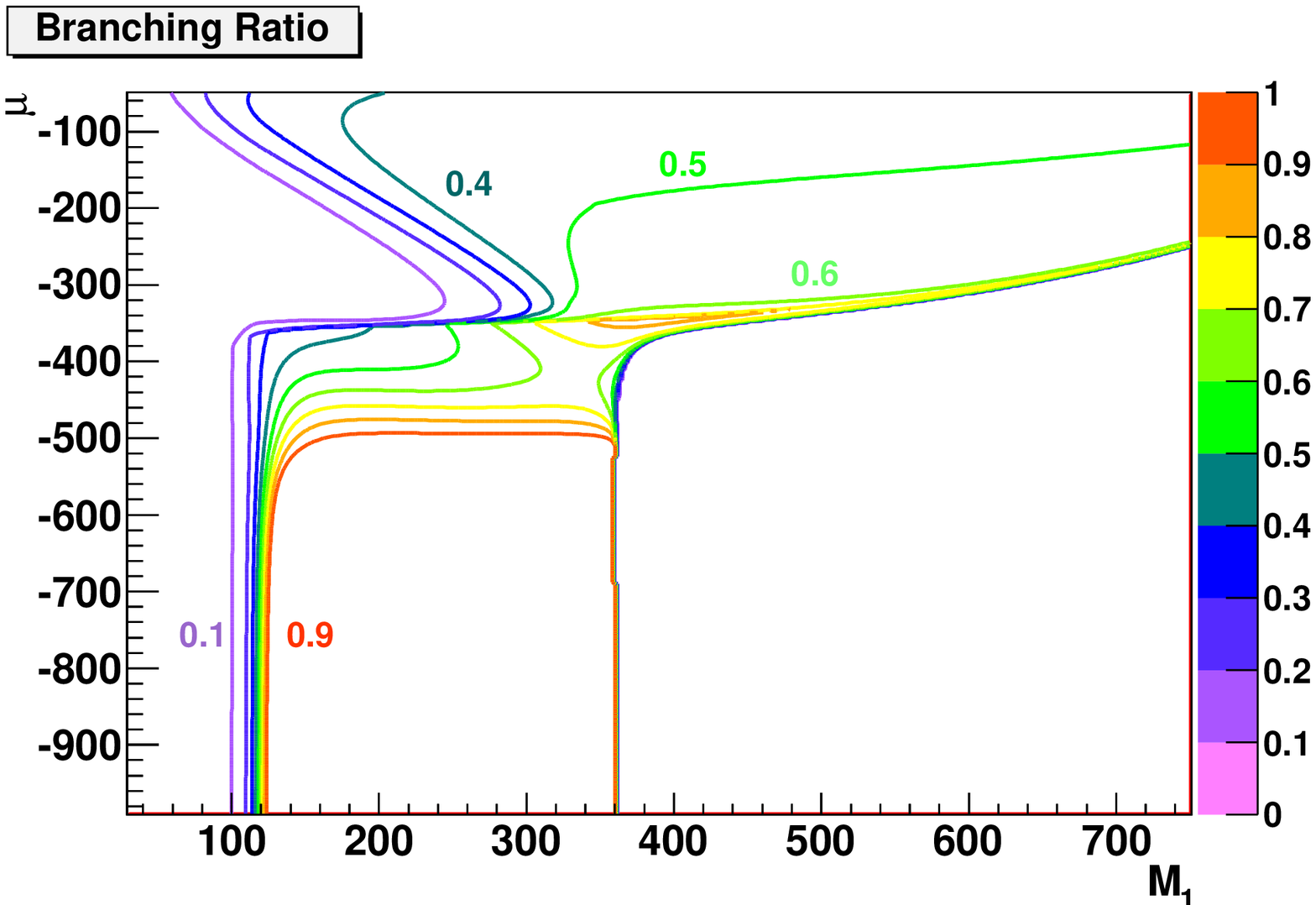}
\caption{Contours of the top polarization in the top rest frame for $\mu <0$ and 
a dominantly LH stop (left panel) with the LH parameters in Tab.~\ref{stop_sec_par}. Branching ratios for $\tilde t_1\rightarrow t \tilde\chi_1^0$  (right panel). In the bottom right corner, the decay is not kinematically accessible.  }\label{fig:contourstopL-}
\end{figure}

\begin{figure}[!h]
\centering
\includegraphics[width=0.49\linewidth]{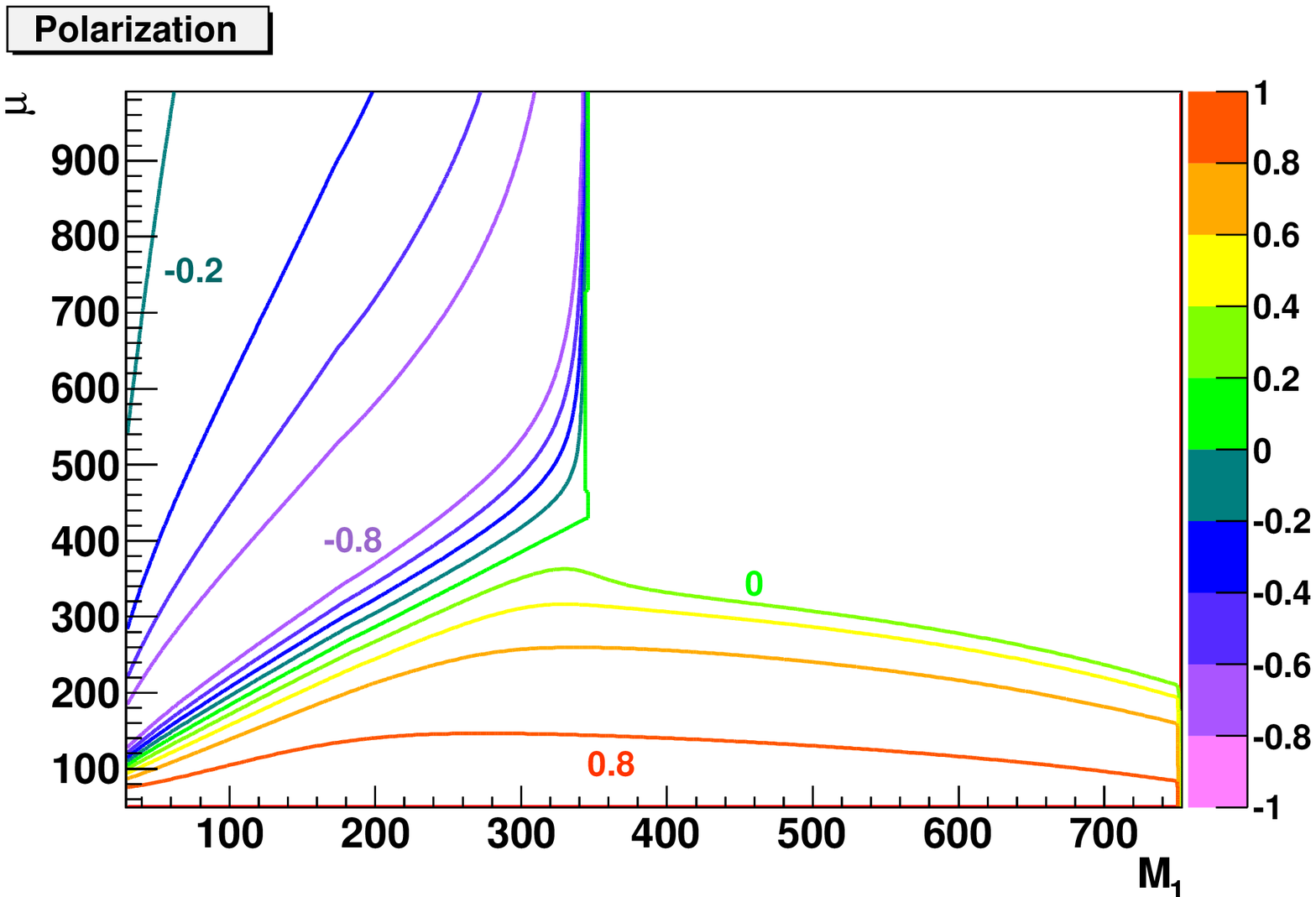}
\includegraphics[width=0.49\linewidth]{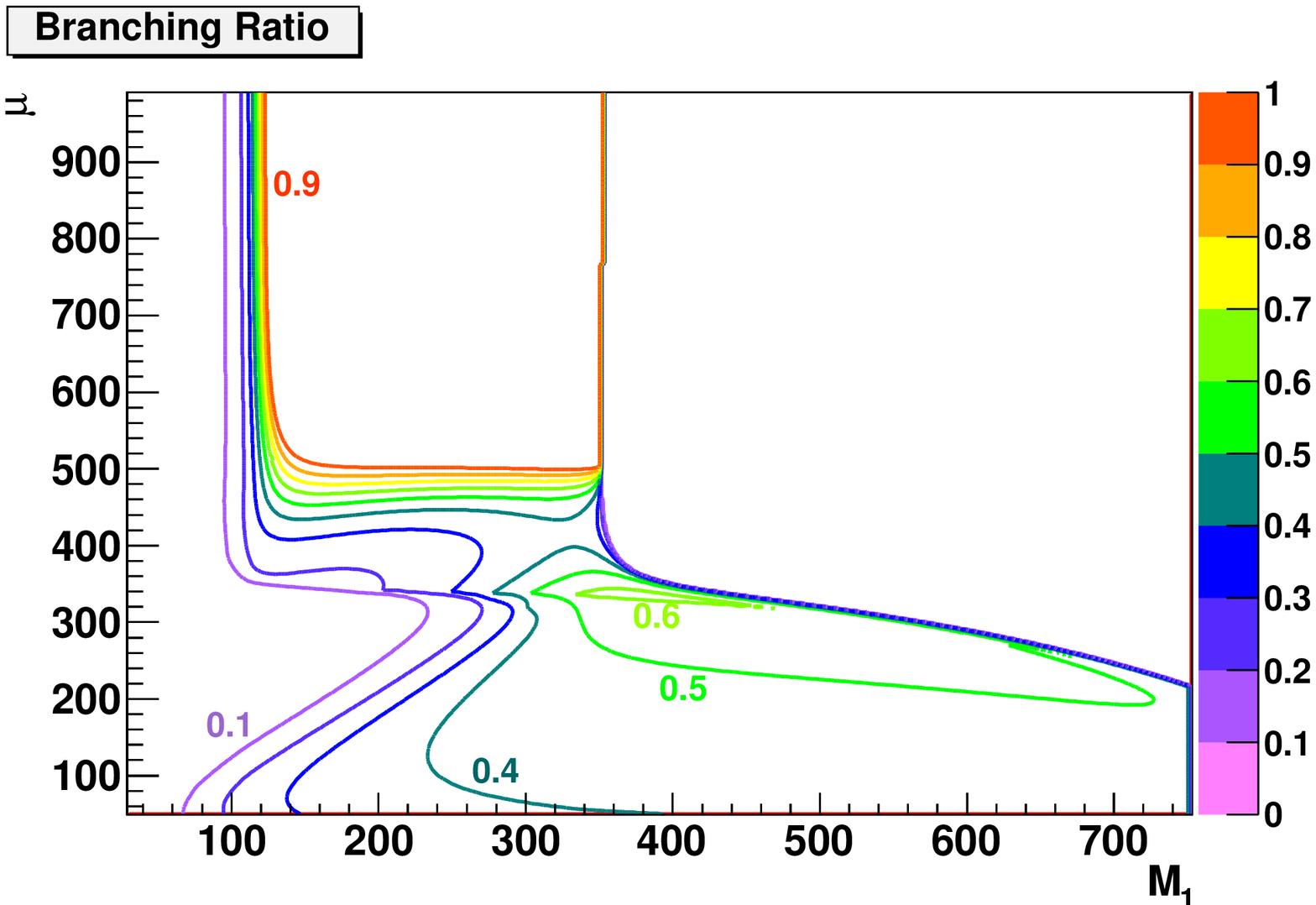}
\caption{Same as Fig.~\ref{fig:contourstopL-} for $\mu >0$ and a mixed but dominantly LH stop
corresponding to the XLH parameters in Tab.~\ref{stop_sec_par}.
In the upper right corner the decay is not kinematically accessible.  }\label{fig:contourstopL+}
\end{figure}

\begin{figure}[!h]
\centering
\includegraphics[width=0.49\linewidth]{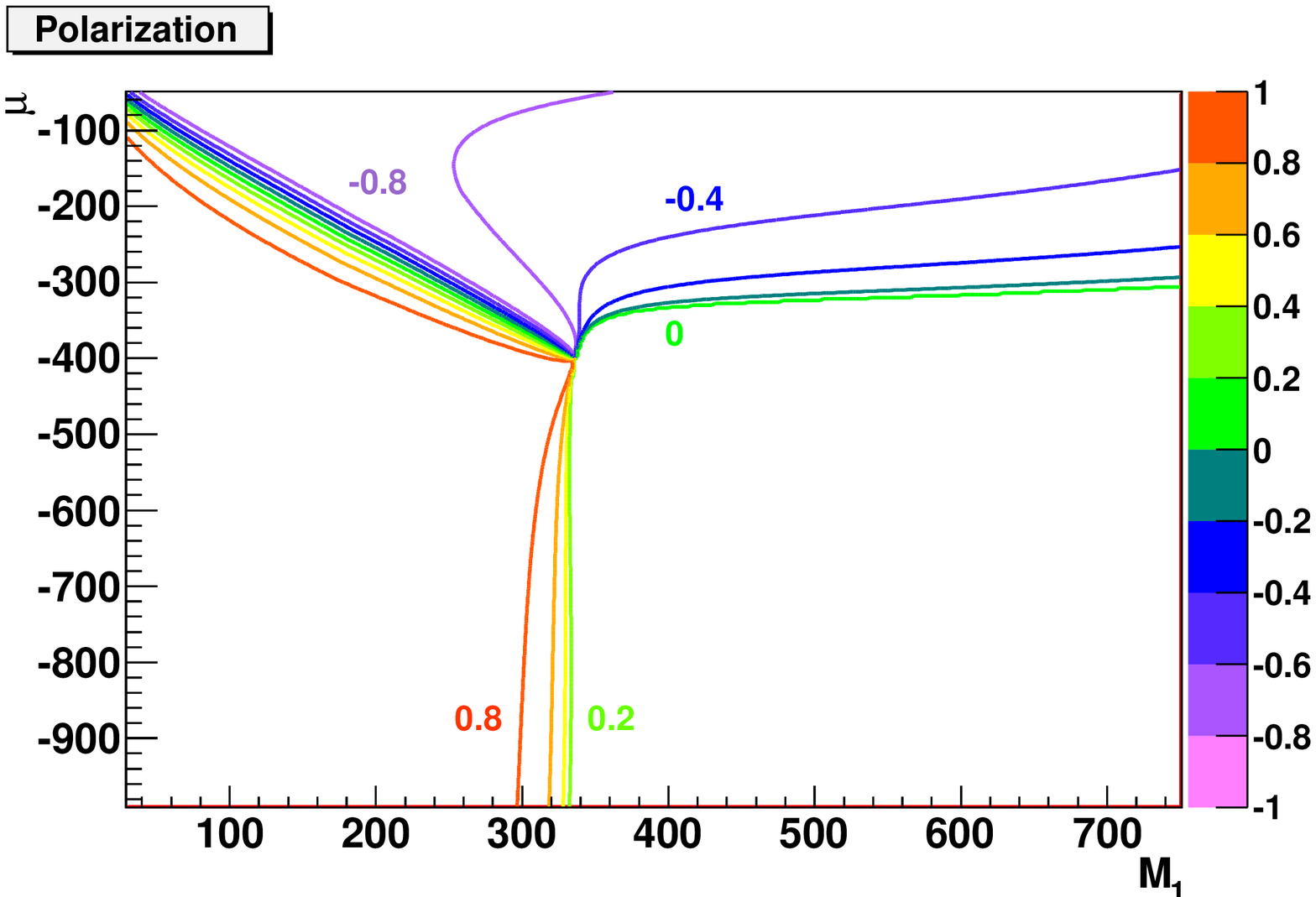}
\includegraphics[width=0.49\linewidth]{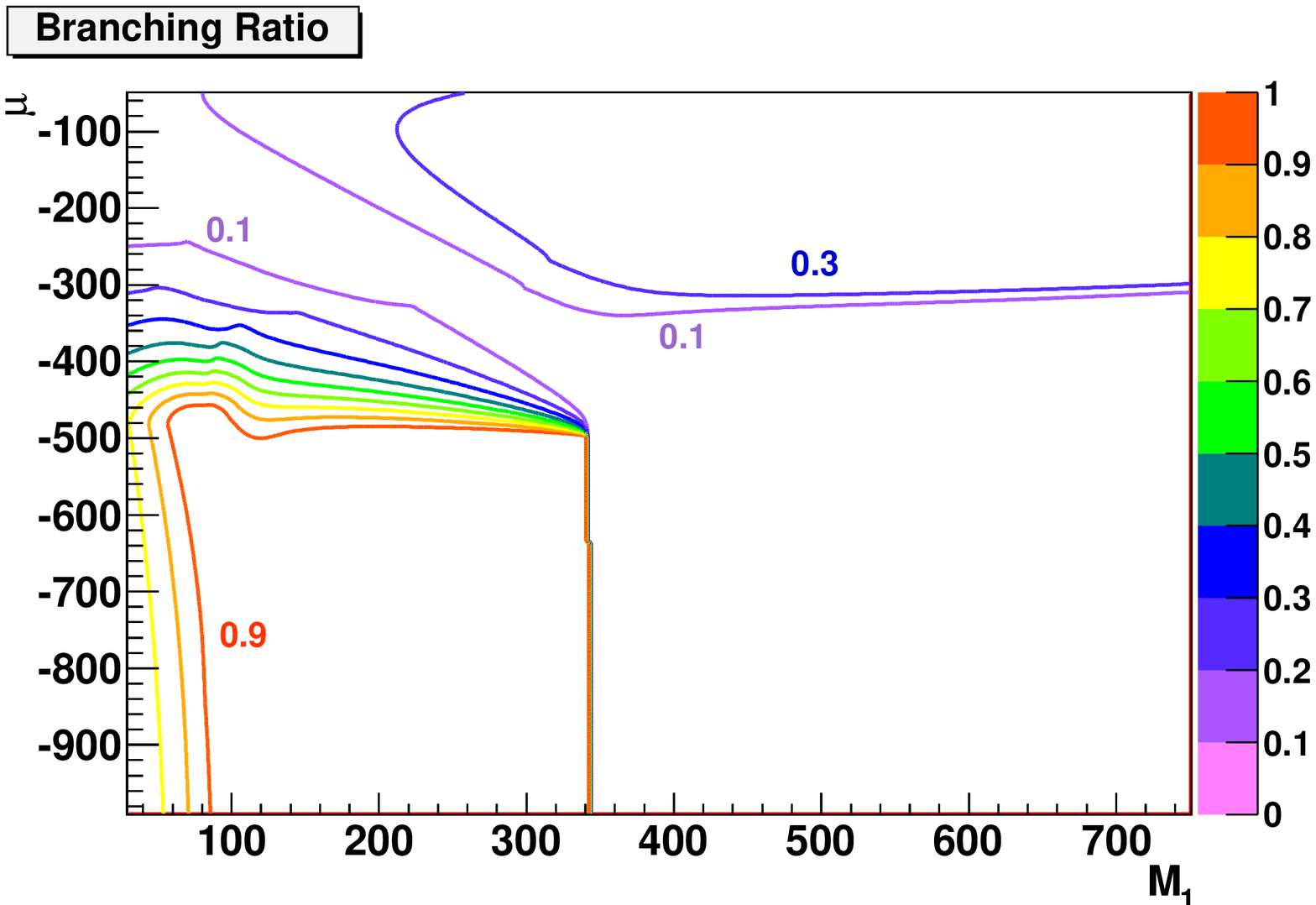}
\caption{Contours of the top polarization in the top rest frame for $\mu <0$ and 
a mixed dominantly RH stop (left). Branching ratios for $\tilde t_1\rightarrow t \tilde\chi_1^0$ (right) 
}\label{fig:contourstopR+}
\end{figure}

\begin{figure}[!h]
\centering
\includegraphics[width=0.49\linewidth]{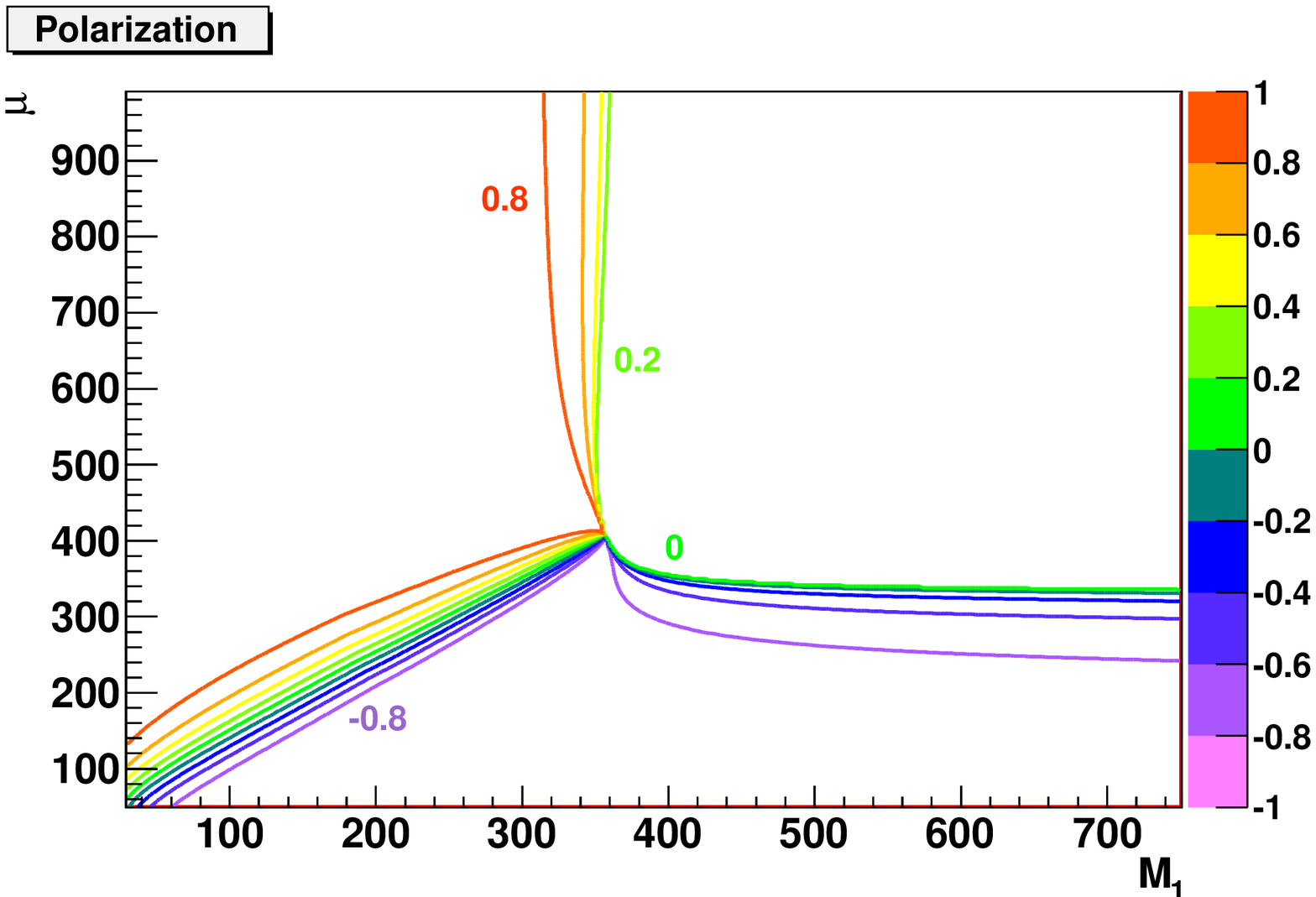}
\includegraphics[width=0.49\linewidth]{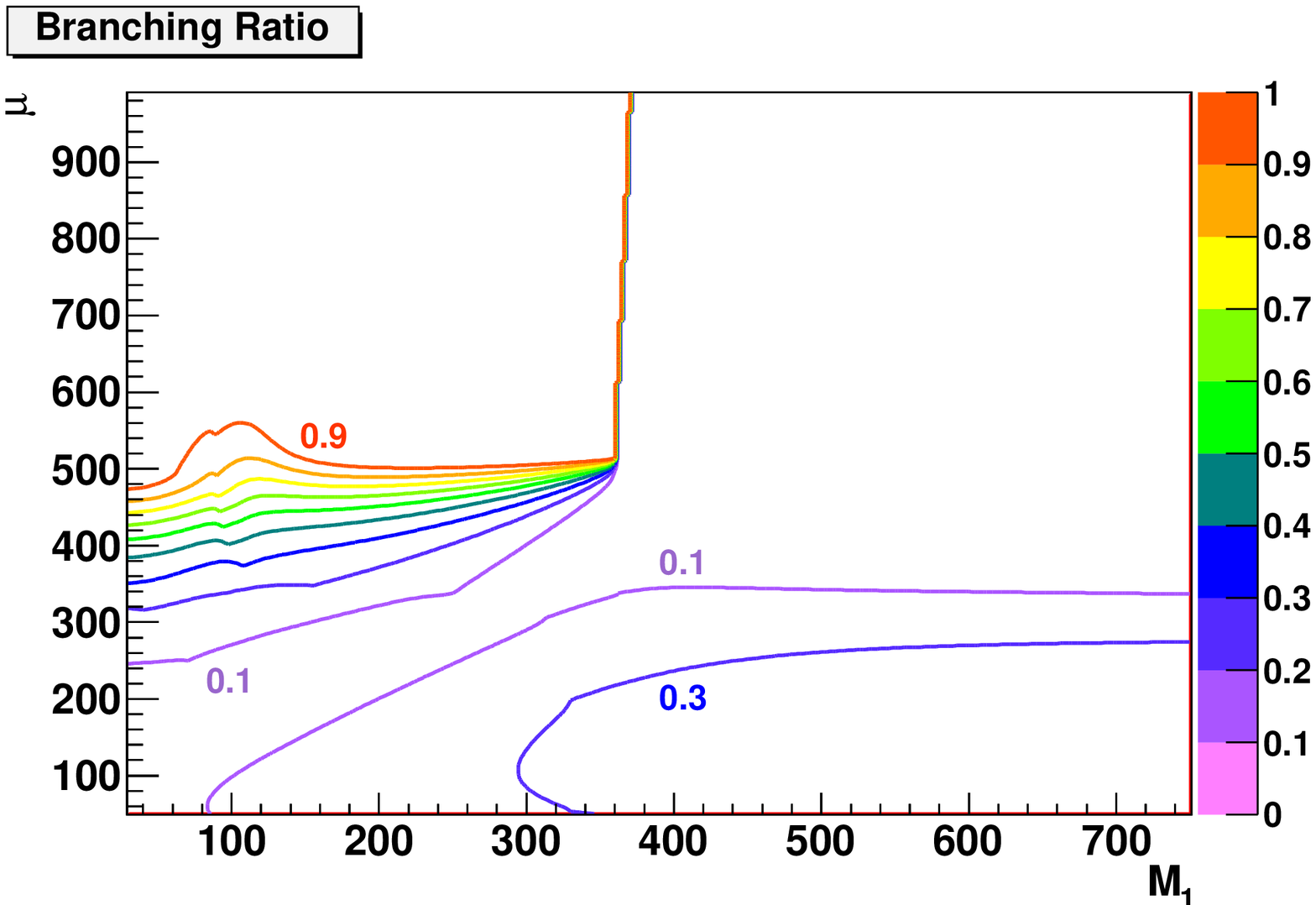}
\caption{Same as Fig.~\ref{fig:contourstopR+} for $\mu > 0$ and dominantly RH stop.  }\label{fig:contourstopR-}
\end{figure}

The contour plots for the top polarization as well as for the  branching ratio $BR(\tilde{t}_1\rightarrow t\tilde{\chi}_1^0)$ in the $\mu-M_1$ planes are displayed in Fig.~\ref{fig:contourstopL-} - \ref{fig:contourstopR-}
for  the four different choices of stop parameters. Here we only consider the region where the decay $\tilde{t}_1\rightarrow t\tilde{\chi}_1^0$ is kinematically accessible. Note that  the maximal variation of the Higgs mass in the  $|\mu|<1$ TeV, $M_1<750$ GeV plane is about 3 GeV, within the theoretical uncertainties, while corrections to $m_{\tilde t_1}$ of the order of  30 GeV can be found for large values of $M_2$ due to the quark/gaugino loop correction.

\vspace{.5cm}
\noindent{\bf The dominantly left-handed stop}
\vspace{.2cm}

As we have discussed in the previous section, in the case of a left-handed stop we expect $P_t \approx -1$  when the LSP is bino-like ($|\mu|\gg M_1$)  and
$P_t\approx 1$ when the LSP is higgsino-like ($|\mu|\ll M_1$).   The polarization contours  in Fig.~\ref{fig:contourstopL-}~(left) for $\mu<0$ illustrate this general behaviour as well as the rapid transition between $P_t=1\rightarrow -1$ in the region where one goes from a bino to a higgsino LSP ($M_1\approx\mu$). 
Note, however, that as the LSP becomes almost pure bino,   the top polarization starts to deviate from $-1$.
For example at $M_1=100~{\rm GeV}, \mu= -600~{\rm GeV}$  the top polarization is only $P_t\approx -0.73$.   
%The  top  polarization in the dominantly bino case   deserves more discussion, as the 
This occurs because we are not dealing with a pure LH stop, indeed here $\sin\theta_{\tilde t}=-0.127$. 
%I find this statement a bit confusing. In which case is what largest? In this case the main contribution to the $G_i^R$ coupling comes from the first term in Eq.~\ref{eq:couplings}, and this term is largest for a pure bino leading to a non-maximal polarization. 
Finally, the kinematic effects which lead to 
$P_t\rightarrow 0$ show up  at the boundary of the grey region. 
 
To be able to exploit the top polarization as an observable, the branching ratio for $\tilde t_1\rightarrow t\tilde{\chi}_1^0$ must be large enough. The contours for this branching ratio are displayed in the right panel of Fig.~\ref{fig:contourstopL-}. 
Large branching ratios are found over most of the parameter space with two exceptions. The first occurs near the kinematic limit where the three-body decay $\tilde t_1\rightarrow b W\tilde \chi_1^0$ dominates and the second occurs for low values of $M_1$. The latter behaviour is a peculiarity due to the fact that we have set $M_2=4 M_1$. Thus for low values of $M_1$ and of $M_2$ the lightest chargino, which is dominantly wino, drops below the mass of the stop and the decay $\tilde t_1\rightarrow b\tilde\chi^+_1$  becomes dominant. If in addition $\mu$ is small, the decay into the second chargino becomes possible as well.

In the region where the LSP is mostly higgsino $|\mu|<M_1$, the mass of the two lightest neutralino and of the lightest chargino are of the same order. Thus the stop can decay  into $t\tilde\chi_1^0,t\tilde\chi_2^0$  as well as into $b\tilde\chi^+_1$. The chargino channel is only at the few percent level while the decay into the LSP increases with the higgsino component reaching a maximum of 70\%. An important fact to keep in mind is that the two lightest neutralinos will have higgsino-components of similar magnitude. Thus the polarization  of the top in the two processes $\tilde t_1\rightarrow t\tilde{\chi}^0_{1,2}$ is similar for the higgsino LSP. Thus one can exploit both decay modes to measure the top polarization, as will be demonstrated below. In the region where the LSP is a bino, $M_1 < |\mu|$, the branching ratio into the LSP is nearly 100\%, except for low values of $\mu$, where the channels $b\tilde\chi^+_1$   (for  $|\mu|<500$~GeV)   and $t\tilde\chi^0_2$ (for $|\mu|<380$~GeV)   also become accessible.

For $\mu>0$, the polarization  and the branching ratio contours have roughly the same behaviour, so we do not  illustrate this case. Rather, we consider a case where the light stop is still dominantly left-handed but where the mixing angle is larger, $\sin\theta_{\tilde t}=-0.223$, see the XLH parameters in Tab.~\ref{stop_sec_par}. The polarization and branching ratio contours are rather similar to the LH case we have just discussed, see Fig.~\ref{fig:contourstopL+}. The main difference is that in the bino region at large $\mu$ and small $M_1$ the polarization is generally not maximal. As we have explained above, the mixing implies that the  main contribution to the $G_i^R$ coupling comes from the first term in Eq.~\ref{eq:couplings},  leading to $|P_t|<1$.  This means that in the bino case, the top polarization is quite sensitive to the mixing in the stop sector.

\vspace{.5cm}
\noindent{\bf The dominantly right-handed stop}
\vspace{.2cm}

Next we consider the case of a dominantly right-handed stop.  
The polarization contours for $\mu<0$ in Fig.~\ref{fig:contourstopR+}
for a mixed RH stop and Fig.~\ref{fig:contourstopR-} for a pure RH stop follow the expected behaviour:  $P_t \approx 1$  when the LSP is bino-like ($|\mu|\gg M_1$)  and
$P_t\approx -1$ when the LSP is higgsino-like ($|\mu|\ll M_1$). 
As before, the kinematic effects (at the boundary of the white region) bring $P_t\rightarrow 0$.  Note also that the sign flip in the polarization as one goes from  the bino/higgsino region is very sharp. The only impact of the larger stop mixing, as illustrated in Fig.~\ref{fig:contourstopR+}, lies in the higgsino region ($\mu<M_1$):  when the mixing in the stop sector is larger, the top polarization is not maximal.
This is because in this case  the main contribution to the $G_i^R$ coupling comes from the second term in Eq.~\ref{eq:couplings},  thus leading to a larger value for $G_i^R$  and $|P_t|<1$.

In both the pure and mixed RH stop cases,  the behaviour of the branching ratio contours are rather similar. The branching ratio $\tilde t_1\rightarrow t\tilde{\chi}_1^0$ is above 90\% in the bino region, except near the kinematic limit where the stop decays only into 3-body, and at low values of $M_1$ for the mixed RH stop. As mentioned above, this is caused by the channel $\tilde t_1 \rightarrow b\tilde\chi^+_1$ becoming kinematically accessible, which is only possible through the LH component of the light stop. In the higgsino LSP region, the BR never becomes very large (up to  roughly 25\% for  $t\tilde\chi_1$ and to 20\% for $\tilde\chi_2, \tilde\chi_3$). Here  the main decay channel is into $b\tilde\chi^+_1$ which has a partial width that is proportional to the top Yukawa coupling for a RH stop and is therefore much larger than in the case of a LH stop where the width is determined by  the bottom Yukawa coupling. Thus for a RH stop and a higgsino LSP, it will be more difficult to measure the top polarization because of the suppressed rate. 

\vspace{.5cm}
\noindent{\bf Decays into heavier neutralinos}
\vspace{.2cm}

\begin{figure}[!h]
\centering
\vspace{-0.5cm}
\includegraphics[width=0.9\linewidth]{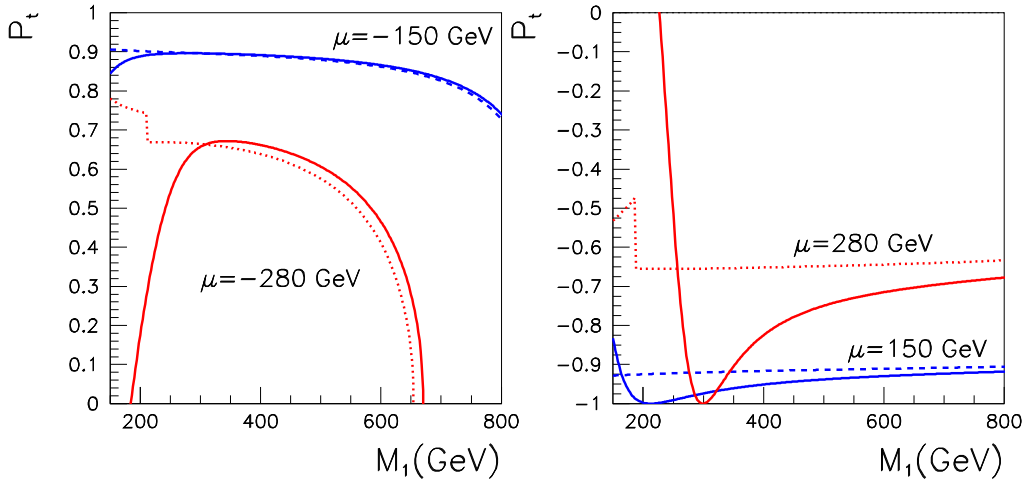}
\vspace{-1.3cm}
\caption{Comparison of the top polarization for the decay $\tilde t_1\rightarrow t \tilde\chi_1^0$ (full line) 
and $\tilde t_1\rightarrow t \tilde\chi_2^0$ (dashed line) as a function of the gaugino mass $M_1$ for  $|\mu|=150$ GeV (blue)
and $|\mu|=300$ GeV (red).
SUSY parameters are fixed  as in Tab.~\ref{stop_sec_par} for a dominantly LH stop (left panel) and 
 a dominantly RH stop (right panel).   }\label{fig:pol2}
\end{figure}

For a higgsino LSP, the branching ratio of the stop into the lightest neutralino can be rather small. However, in this case the top polarization is almost the same when one considers the decay $\tilde{t}_1\rightarrow t\tilde\chi^0_{1}$ or
$t\tilde\chi^0_{2}$ as illustrated in Fig.~\ref{fig:pol2}. 
For the  dominantly LH stop (left panel), the difference between the polarizations in the two channels never exceeds 10\% when $M_1>\mu$ which marks the unset of the higgsino LSP region.  For the  RH stop (right panel) the difference between the polarizations can reach 30\% when $M_1\approx \mu=280{\rm GeV}$   although both polarizations quickly become almost equal as $M_1$ is increased and thus the higgsino fraction of the neutralinos. The difference between the top polarization in the two higgsino channels  is purely a kinematic effect due to the smaller mass splitting between the stop and the second neutralino. This effect is more pronounced for the RH stop case simply because the mass of $\tilde{t}_1$ is lower.
Note that since the two lightest neutralinos are almost degenerate the decay of the second neutralino into the LSP is accompanied by soft leptons and has basically the same missing $E_T$ signature as the LSP. One can therefore use both decay channels to determine the top polarization without being handicapped by small rates. 

In the above, we have considered only the behaviour of the top polarization without worrying about other constraints on the model. We briefly comment on the impact of these constraints.
For the bino case the relic density is typically much too large, it is however possible to bring it to a reasonable value by decreasing  the mass of the sleptons to just above the LSP mass thus adding an important contribution from coannihilation processes. This would have no impact on the polarization observables discussed here. In the higgsino region, as expected the relic density is typically too small. This only means that the neutralino cannot form all of the dark  matter. Constraints on observables from the flavour sector are easily satisfied. For instance the branching ratio for $B_s\rightarrow \mu^+\mu^-$ remains near the SM value since we are considering only moderate values of $\tan\beta$ and a heavy pseudoscalar. For the same reason, the value for  $b\rightarrow s\gamma$ falls within the allowed range in the dominantly RH stop case where we have fixed a high mass for the sbottom. However, this observable can receive too large contributions from sbottom/gluino corrections in  the LH stop scenario since the LH sbottom is also not too heavy. These contributions can be cancelled, bringing the value for $b\rightarrow s\gamma$ back within the measured range by adjusting the pseudoscalar mass.

\section{Top Polarization: Effect  on decay kinematics and observables}\label{s:observables}

We have seen in the previous sections how the top polarization is influenced by (1) the mixing of the stop and neutralinos and (2) the masses of the particles in the decay chain. In this section, 
we first study the effect of this polarization of the decaying top 
on the kinematics of the lepton produced in its 
semi-leptonic decay (Eq.~\ref{eq:topdecay}) and assess the possible 
effects top polarization can have for the search strategies for 
the stop. Further we  study qualitatively if top polarization at the LHC, measured via this semi-leptonic decay  
can be a useful probe for the neutralino and stop mixing parameters when there is prior knowledge on SUSY masses. We start by reviewing the decay of the top. We will also see that the angular observables of the semi-leptonic decay can provide a pure measure of  polarization.

The polar angle distribution of the top decay product $f$ is described, in the top rest frame, by 
\begin{equation}
\frac{1}{\Gamma_l}\frac{\mathrm{d}\Gamma_l}{\mathrm{d}\cos\theta_{f,\rm rest}}=\frac{1}{2}\left(1+ \kappa_f P_t\cos\theta_{f,\rm rest}\right),\label{eq:topdecayeq}
\end{equation}
with $\Gamma_l$ the partial decay width, $\theta_{f,\rm rest}$ the rest frame angle between
decay product $f$ and the top spin vector, $\kappa_f$ the analyzing
power of the decay product and $P_t$ the polarization of the top. 
Effects of polarization are studied most easily for a decay to a positively
charged lepton or a down-type quark in which case $\kappa_f=1$. The value of
$\kappa_f$ is only mildly influenced by higher order corrections and non standard
$tbW$ couplings. 
%gb The former slightly reduce $\kappa_f$  at the permille level 
The former induce corrections to  $\kappa_f$  at the permil level for a decay to a down-type quark~\cite{Brandenburg:2002xr},
whereas the latter do not influence  $\kappa_f$ at leading
order~\cite{Godbole:2010kr}. Therefore the leptonic decay provides a
good probe for the polarization of the top quark, even in the presence
of such anomalous couplings. We will further only consider top quarks,
since the anti-top can be distinguished by the charge of the decay
lepton. In fact while measuring the  polarization, one can double the sample by using decays of both the tops and the anti-tops.

One obvious way to measure the polarization of the top is to construct the rest frame of the decaying top. We will here 
look here however, at the laboratory frame observables with 
a two fold objective. This will give us an idea of the effect that the top polarization can have on the kinematics of the decay 
lepton in the laboratory frame  and hence on search strategy. Further,  it may  not be necessarily easy to construct the 
rest frame of the top at LHC and also because observables constructed out of the laboratory variables can provide an alternate measure of the top polarization. 

The use of laboratory frame means that the polar distribution $\theta_l$ of the top  decay products is now 
described by Eq.~(\ref{eq:topdecayeq}) and 
the subsequent boost from the rest frame to the lab frame. 
The
azimuthal distribution, which is uniform in the rest frame, is
influenced by the kinematics of the stop production process through the boost. 
To determine the azimuthal angle $\phi_l$, we must define a
frame. The $z$ axis is taken to be the beam direction, and the
direction of top momentum together with the beam axis defines the $xz$
plane. The $y$-axis can then be constructed according to the
right-hand rule. 

To examine the effect of the top polarization on  the kinematic distributions of the semi-leptonic top quark decay product we have generated sets of events with Madgraph
~\cite{Alwall:2011uj,Alwall:2007st}. This set of benchmarks has been
selected based on the degree of top polarization in the stop rest frame as well as a 
%gb the mass difference is not exactly constant so I modified here and in the table caption later
roughly constant mass
difference between stop and neutralino. The physical parameters corresponding to these
benchmarks are listed in Tab.~\ref{bench2}. We have generated the process 
\begin{equation}
p\; p \rightarrow {\tilde t} \; \bar{{\tilde t }} \rightarrow t \,{\tilde \chi}_0 \, \bar{{\tilde t }} \rightarrow l^{+}\,\nu_l\, b \, {\tilde \chi}_0 \, \bar{{\tilde t }} 
\label{eq:stopantistop}
\end{equation}
We took $8$ TeV as LHC center of mass energy  
and use the following
parameter values: the top mass and width are $m_t=173.1$ GeV and
$\Gamma_t=1.50$ GeV, and the $W$ mass and width are $m_W=79.82$ GeV and
$\Gamma_W=2.0$ GeV. The factorization and renormalization scales are set
to $\mu_R=\mu_F=m_{\tilde t}$. It was shown in \cite{Godbole:2011vw} that NLO corrections do not change the qualitative features of the lab-frame observables constructed out of the angular variables, so we show leading-order (LO) results, which were calculated with the CTEQ6L1 \cite{Nadolsky:2008zw} pdf set. 

\begin{table}[!h]
\centering
\begin{tabular}{c|ccccccc}
$P_t$ & $m_{\tilde t}$ (GeV) & $m_{\tilde \chi^0_1}$ (GeV) &  $\sin(\theta_{\tilde t})$ & $Z_{i1}$ & $Z_{i4}$ & $\tan(\beta)$ \\ 
\hline
 1 & 500.0 & 318.6 &         0.998   & 0.958 & -0.176 & 7.8 \\
 0.5 & 500.0 & 321.1 & 0.998 & 0.988 & -0.0866 & 7.8 \\
  0 &  500.0 & 320.5 &      -0.124  & 0.975 & -0.128 & 10.0 \\
 -0.5 & 501.1 & 319.2 &         0.995 & 0.440 & -0.618 & 20.0 \\
 -0.8 & 502.0 & 319.3 &        -0.0988 & 0.0232 & -0.190 & 35.0 \\ 
\hline
 1 & 500.7 & 130.2 &        0.9928   & 0.9976 & -0.1883 & 10. \\
 0.5 & 499.6 & 129.7 &  0.9987 & 0.9164 & -0.2112 & 29.6 \\
0 & 500.1 & 129.3 &  -0.05954 & 0.9729 & -0.1017 & 35.0 \\
 -0.5 & 500.1 & 130.3 &        -0.05948 & 0.9865 & -0.06113 & 35.0 \\
 -1 & 499.4 & 130.0 &       -0.05911 & 0.9990 & -0.007184 & 35.0 
\end{tabular}
\caption{Set of benchmarks sorted by polarization. The upper five correspond to small mass differences and the lower five to large mass differences. The mass of the second neutralino is shown for the cases where its branching is non-zero.}
\label{bench2}
\end{table}

\subsection{Effect of top polarization on $E_{l}$ and $P_{T}^{l}$}
In this subsection we show the effect of the top polarization on the energy $E_{l}$ and the transverse momentum $P_{T}^{l}$  of the lepton  produced in the decay of the top in the laboratory frame for our benchmark points.  These two distributions in the laboratory 
depend on the  angular distribution of the lepton given in 
Eq.~\ref{eq:topdecayeq} in the top rest frame, as well as  the energy and the $P_{T}$ of the decaying top which decides the direction and the magnitude of the boost to the laboratory frame. Since the angular distribution of Eq.~\ref{eq:topdecayeq} depends on the polarization of the decaying top, the $E_{l}$ and $P_{T}^{l}$ distributions have a dependence on the top polarization. Most of the decay leptons in the rest frame come in the forward direction for a positively polarized $t$ quark , i.e. the direction of the would-be momentum of the $t$ quark in the laboratory. Thus
after a boost from the rest frame to the lab frame the 
energies of these leptons are increased.  Similarly, for negative polarized $t$ quarks most of the decay leptons come out in the
backward direction w.r.t. the lab momentum of the $t$ quark. 
This results in an opposite boost direction  and hence a 
decrease in the energy of the leptons. The effect on the $P_{T}$ distribution of the lepton in the laboratory is further also affected by the $P_{T}$  of the $t$ quark as well.

\begin{figure}[!h]
\centering
\includegraphics[angle=0,width=0.49\linewidth,]{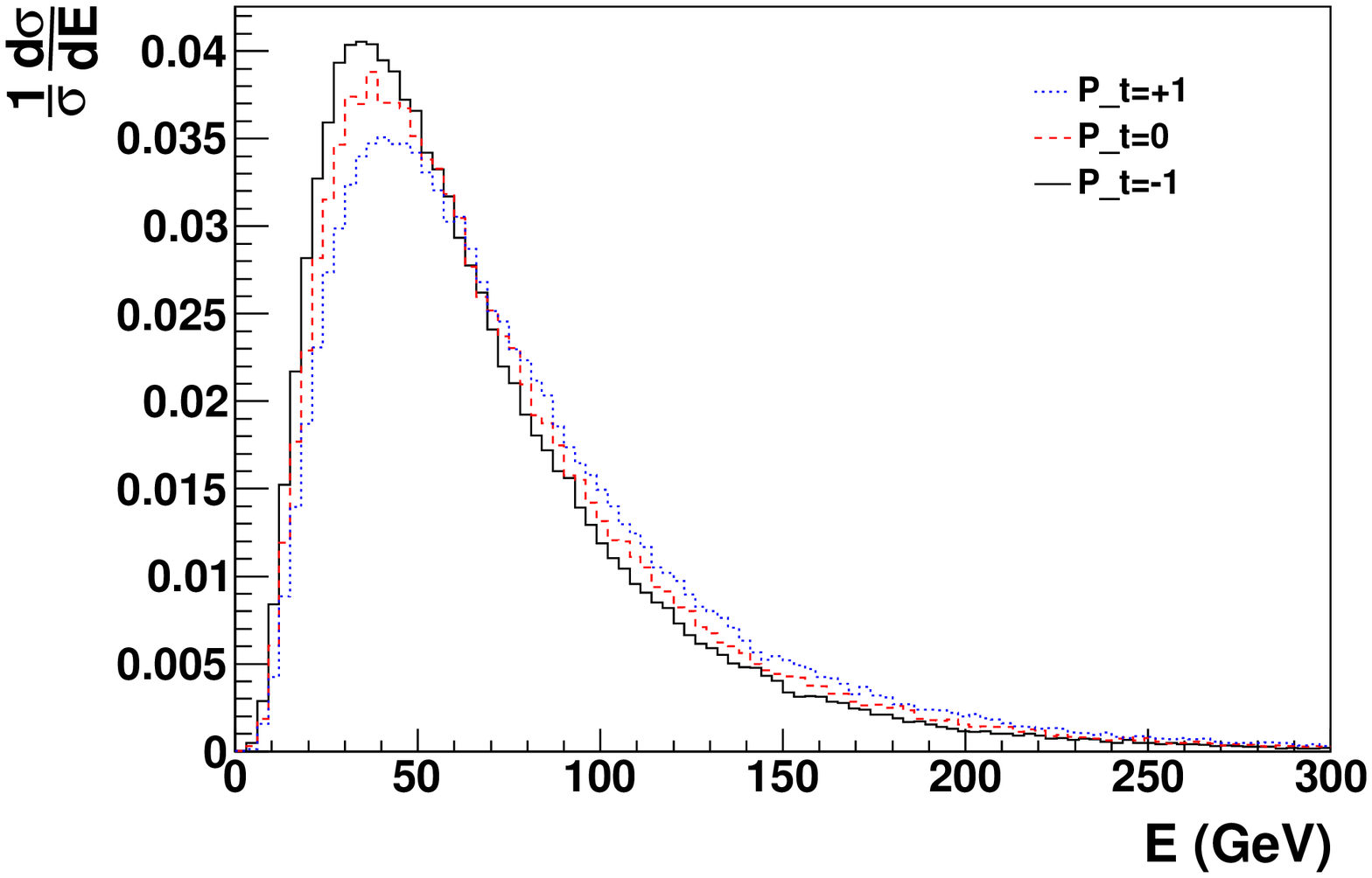}
\includegraphics[angle=0,width=0.49\linewidth,]{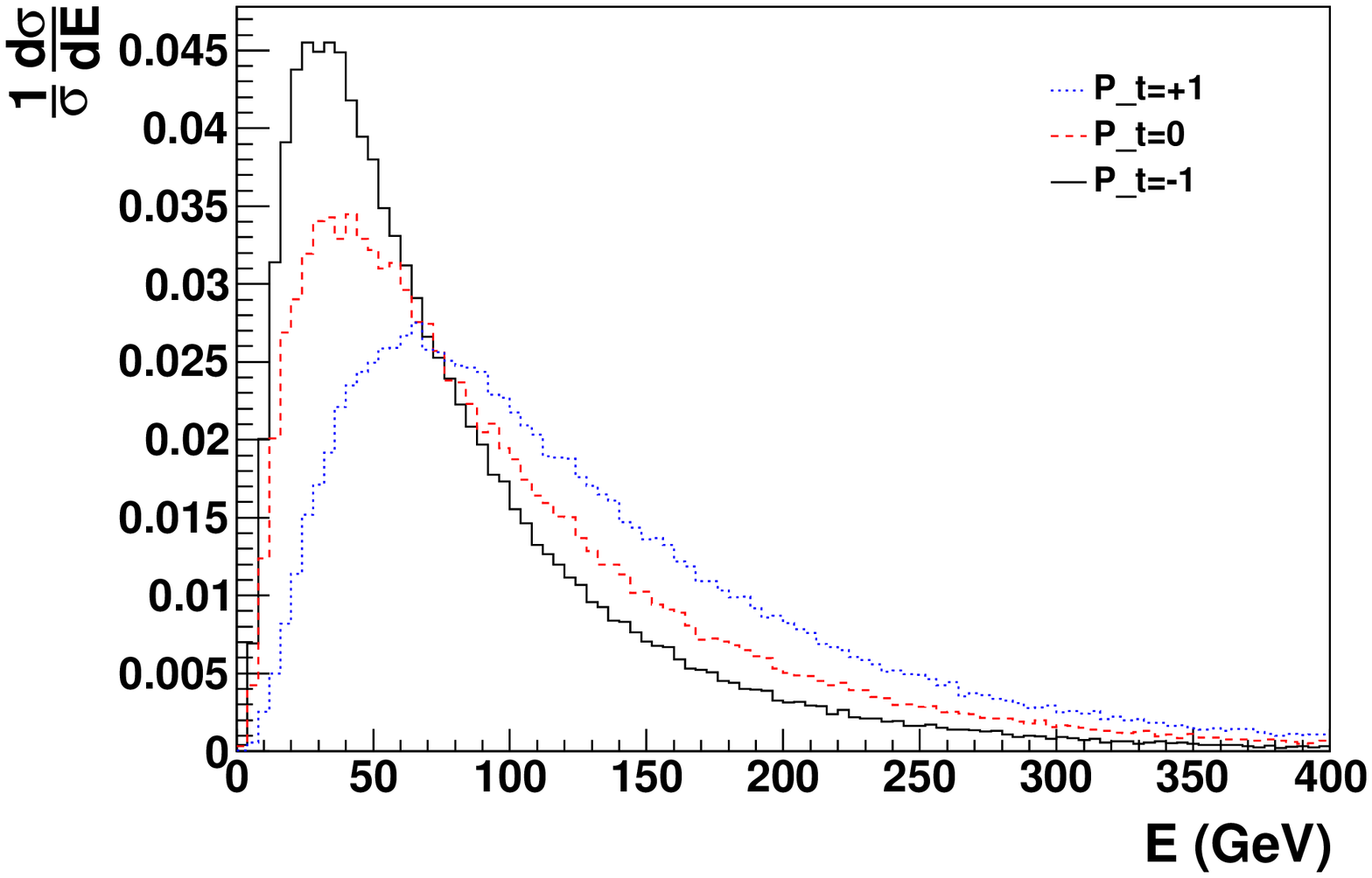}
\caption{The  distribution in the energy  of the lepton coming from
 the decay of the top quark, for three different polarizations of
 the decaying $t$ quark: 1,0 and -1 being given by the 
 blue, red and the black lines respectively.. The left graph represents benchmarks with a small mass difference and the right graph benchmarks with a large mass difference between stop and neutralino.}\label{fig:El_dist}
\end{figure}

\begin{figure}[!h]
\centering
\includegraphics[angle=0,width=0.49\linewidth,]{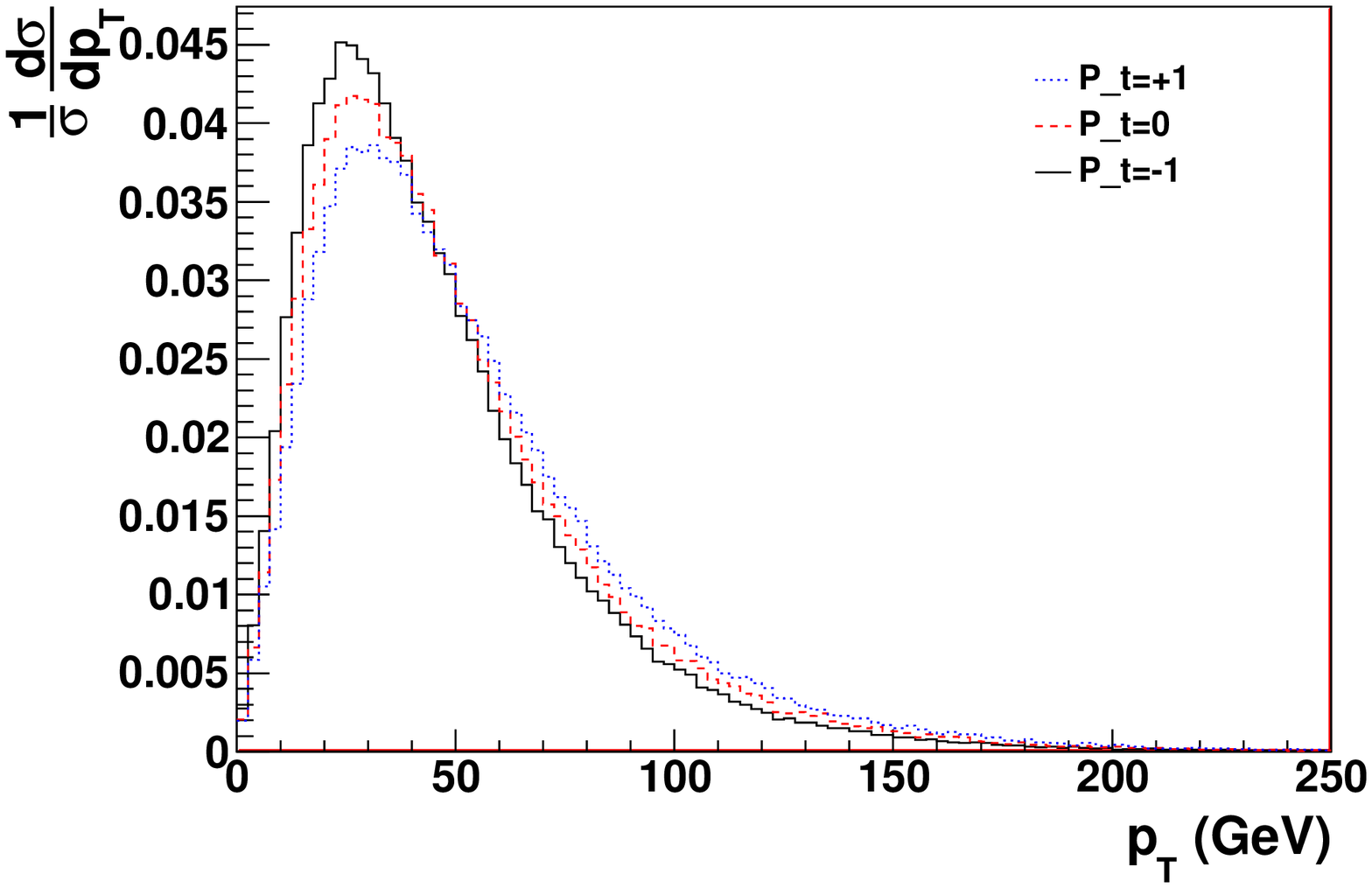}
\includegraphics[angle=0,width=0.49\linewidth,]{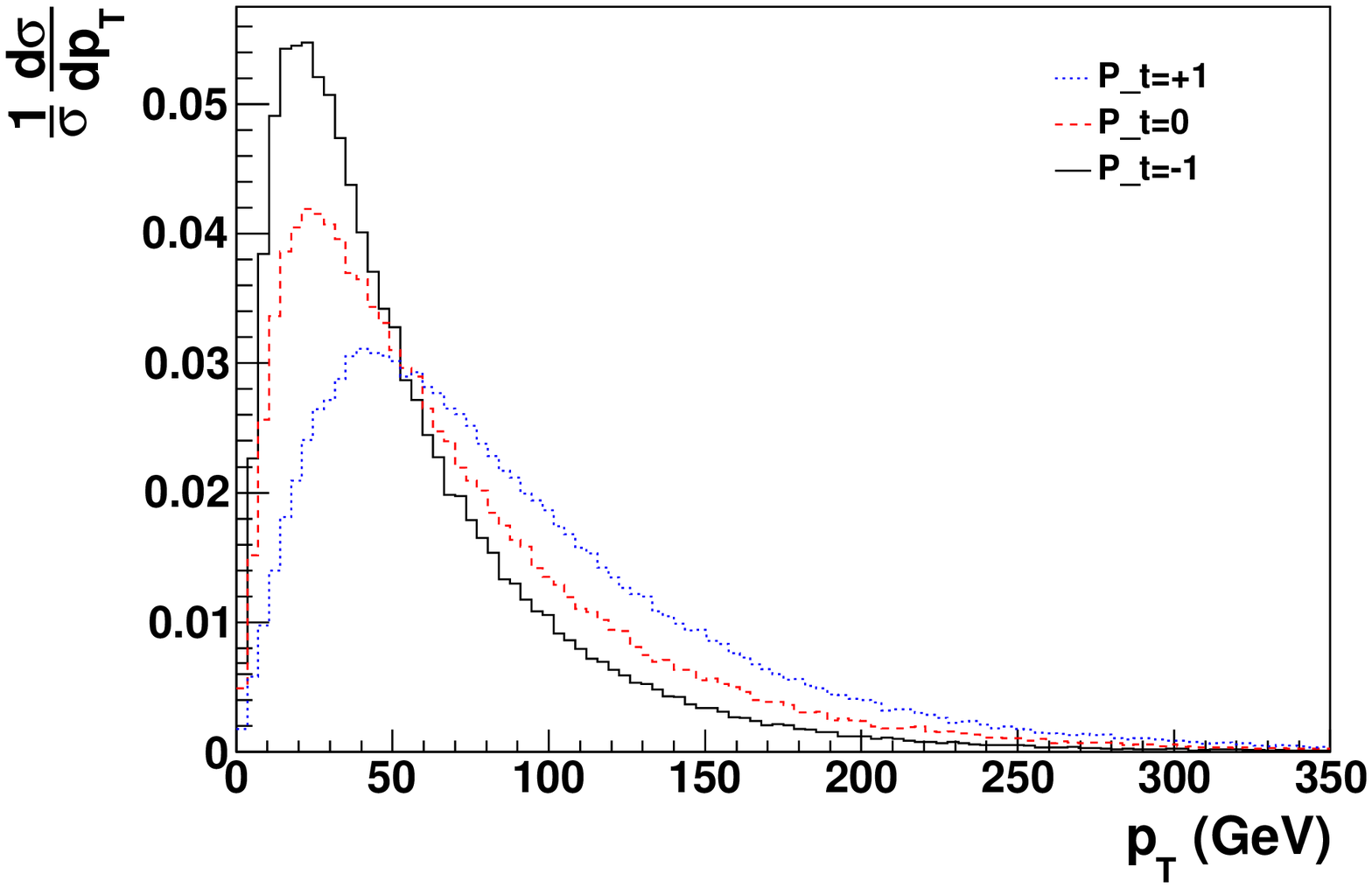}
\caption{The distribution in $P_T$ of the  lepton coming from the decay of the top quark. The left graph represents benchmarks with a small mass difference and the right graph benchmarks with a large mass differences between stop and neutralino.}\label{fig:pt_dist}
\end{figure}

Fig.~\ref{fig:El_dist} shows the $E_{l}$ distribution in the laboratory for three different polarizations of the parent top quark: $1$, $0$ and $-1$, being depicted in blue, red and black respectively. 
%Since, for  the three cases, in each figure, the   kinematics of the produced stop and the top is the same, 
%the difference in {\it mass differences and stop masses} being negligible to cause  any effect, the entire difference is due to  polarization of the decay top. 
Since, for  the three cases in each figure, the mass difference between the stop and the top is nearly the same,  
the entire difference in the distributions can only be due to  polarization of the decay top. 
Consistent with the qualitative argument given above,  the peak of the $E_{l}$ distribution shifts  to lower energies for the left polarized top with respect to an unpolarized top and  to higher energies for the right polarized one. The shift is higher for the case of large mass differences (with peaks occurring at respectively $26$, $42$ and $66$ GeV) compared to the small mass difference (with peaks occuring at $34.5$, $37.5$ and $40.5$ GeV). Since, one puts cuts on the lepton kinematic variables to reduce the background from the SM tops (which would have polarization zero) one sees that such cuts will 
be less effective for a left polarized top and it will be even more so for the case of large mass differences. The distributions
for the transverse momentum of the lepton, shown in 
Fig.~\ref{fig:pt_dist} shows similar features. For small mass differences the transverse momentum distribution of a polarization of $-1$, $0$ and $+1$ respectively peaks at $24$, $26$ and $31$ GeV. For large mass differences the distribution of a polarization of $-1$, $0$ and $+1$ respectively peaks at $26$, $42$ and $66$ GeV.
In fact we also notice that 
the shifts in the $P_{T}^{l}$ distributions are substantial compared to the possible effects which would come from changes in the $P_{T}^{\tilde t}$ distribution coming from NLO 
effects~\cite{Beenakker:1998wi,Beenakker:1997ut,Beenakker:2010nq}
So, this effect needs to be taken into account even in an analysis that neglects the NLO effects on the stop production.

Thus we clearly see that the current limits quoted on the stop quark mass from direct production, using the 
$t \tilde \chi_{1}^{0}$ channel, will depend on the amount of top polarization and in addition the effect of the mass difference $m_{t}- m_{{\tilde \chi_{1}^{0}}}$. This needs to be kept in mind while assessing the limits being quoted currently. The observation above also means that the searches for the stop with SUSY parameters, which give rise to negatively polarized tops are in fact doubly challenged as the single top background will also produce top quarks which are negatively polarized. Whereas for the case of positively polarized top quarks being produced by SUSY, one can use the above distribution to discriminate effectively against the background coming from single top quark production.

This also means that, in principle, information on the energy of the lepton may be used as a `measure' of the parent top
polarization. In fact, for heavily boosted top quarks, studying distributions in fractional energy of the  decay lepton and $b$ quark has been shown to carry information about the top polarization~\cite{Shelton:2008nq}. In fact a recent study
demonstrates their use  for the case of hadronically decaying tops, at the $14$ TeV LHC
\cite{Bhattacherjee:2012ir}.
It should be noted, however, as mentioned earlier,
that the energy distributions  of the decay products can be affected by
the anomalous $tbW$ coupling and hence are less robust a measure of the top polarization of the parent top quark, than the angular 
observables~\cite{Godbole:2006tq,Godbole:2011vw}. We discuss these 
in the next subsection. 
%\todo{after this I have to add few lines about why one does not use these distributions for determination of polarization}
   
%gb here I added a subsection to make it easier for the reader to follow the structure of the section
\subsection{Observables}
In this subsection now we focus on the observables which
will give us a measure of the polarization of the top quark, using  angular observables of the decay lepton which has the highest 
analyzing power, which is furthermore unaffected by the anomalous $tbW$ coupling to the leading 
order~\cite{Grzadkowski:1999iq,Grzadkowski:2002gt,Grzadkowski:2001tq,Hioki:2002vg,Ohkuma:2002iv,Rindani:2000jg,Godbole:2002qu}.
We explore utility of various asymmetries constructed out of the $\phi_{l}$ and $\theta_{l}$ distributions, as in 
~\cite{Godbole:2002qu,Huitu:2010ad,Rindani:2011pk,Godbole:2011vw}.  
%We do not explore the observables constructed out of distributions
%in the variables $u,z$, energy fractions of the decay lepton 
%and the $b$ quark of Ref.~\cite{Shelton:2008nq} in this study.

%\todo{Now here I will add a few lines which will connect with the %last lines in the above subsection and say that in this subsection %we will discuss the observables. That will be a line or two.}
 
\vspace{.5cm}
\noindent{\bf Azimuthal asymmetries}
\vspace{.2cm}

\begin{figure}[!h]
\centering
\includegraphics[angle=0,width=0.49\linewidth,]{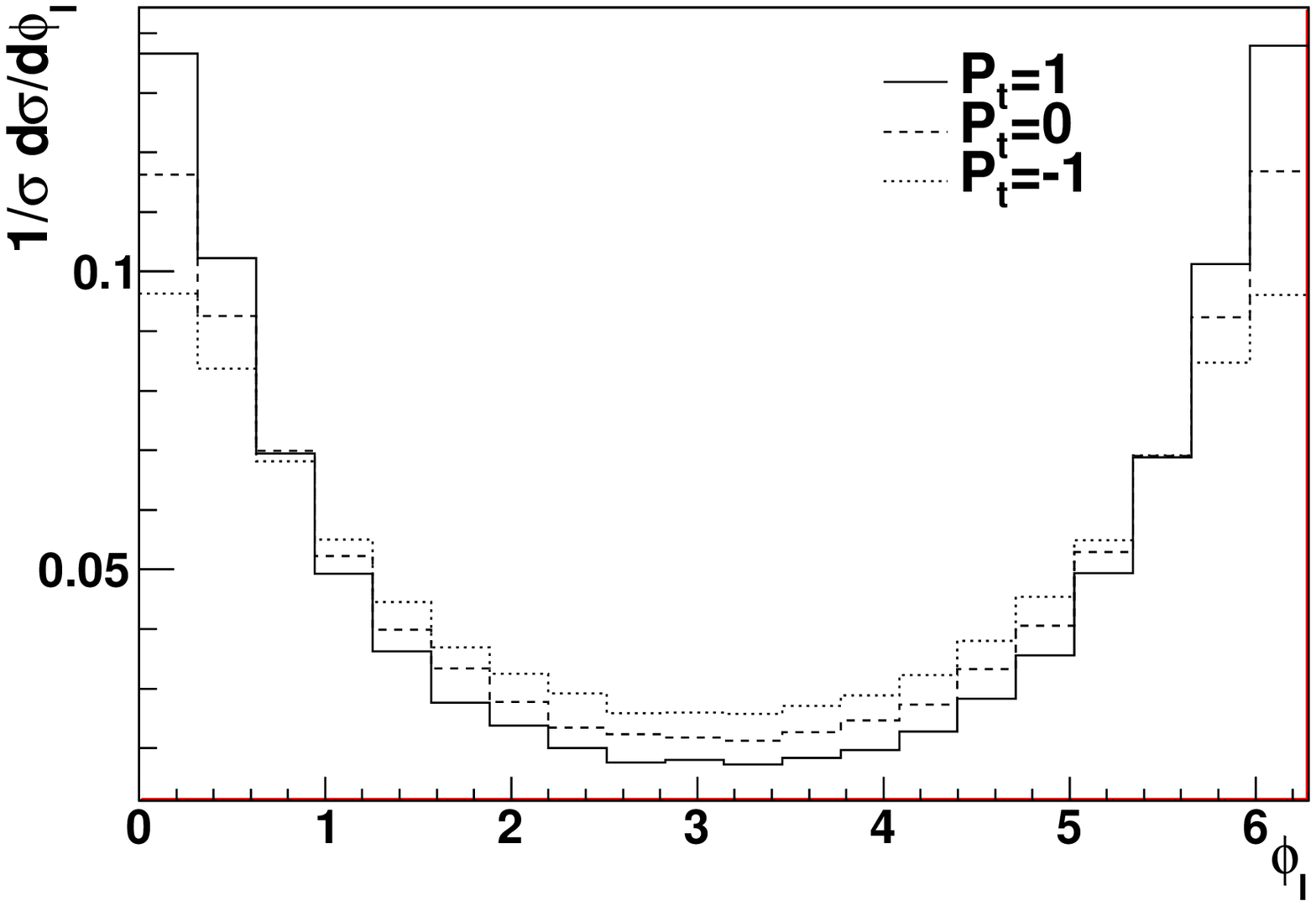}
\includegraphics[angle=0,width=0.49\linewidth,]{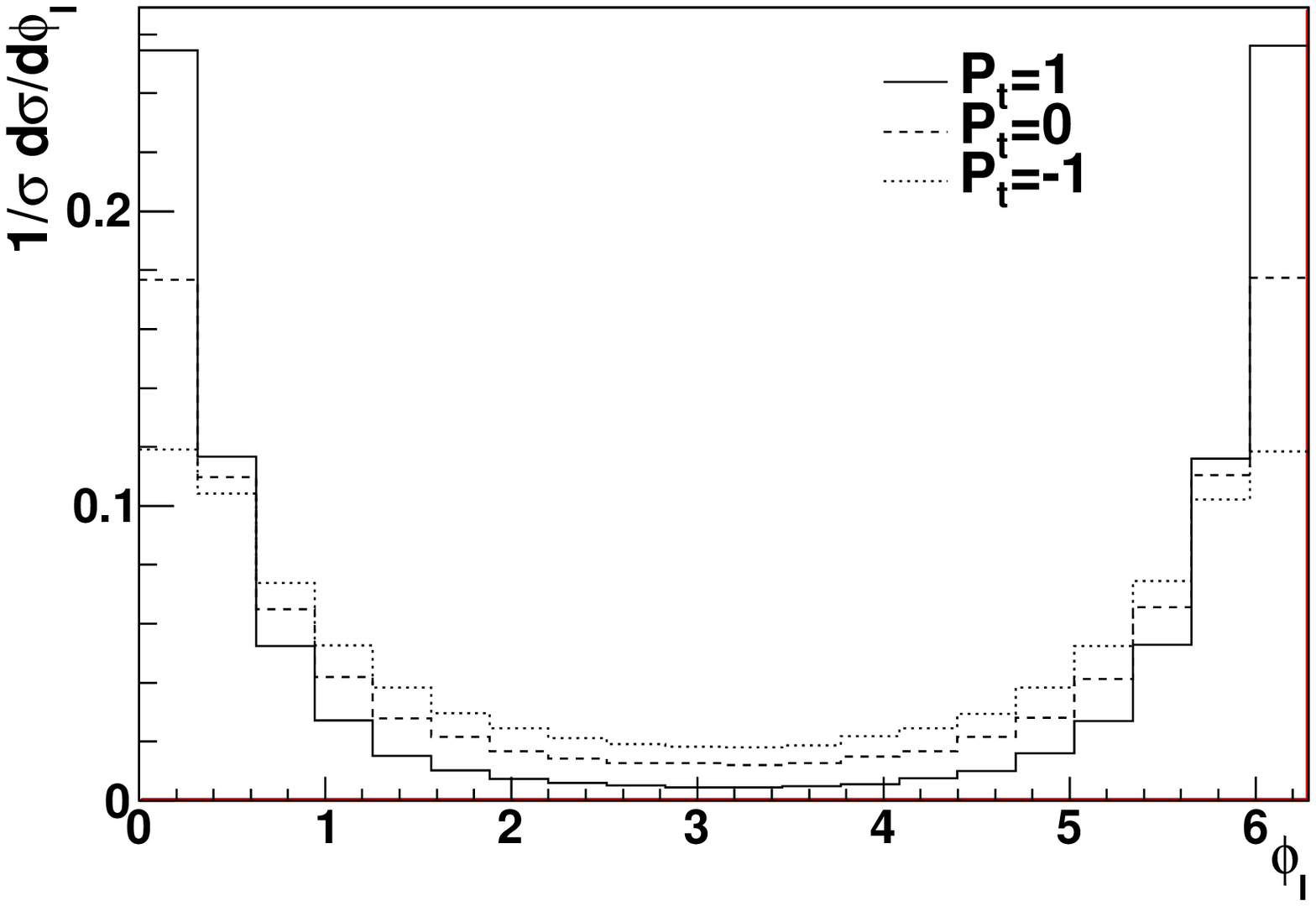}
\caption{The azimuthal distribution $\phi_l$ of the decay lepton of the top quark. The left graph represents benchmarks with a small mass difference and the right graph benchmarks with a large mass differences between stop and neutralino.}\label{fig:Phi_dist}
\end{figure}

The azimuthal distributions of the charged lepton from top decay
for selected benchmarks are plotted in
Fig.~\ref{fig:Phi_dist}. The left plot contains the benchmarks with a
small mass difference between stop and neutralino, and the right
plot those with a large mass difference. The
distributions peak at $0$ (and of course $2\pi$), with the stronger peaking for
a positively polarized top. The unpolarized top case ($P_t=0$ benchmarks) 
%gb exposes the influence of the kinematics, 
illustrates the  influence of the kinematics, 
since an unpolarized top generates a uniform
distribution of decay products in the rest frame. The boost gathers the decay products
towards the boost axis. The boost axis in the
$xy$-plane coincides with the $x$-axis, which is defined by the top momentum in
this plane, so around this axis all distributions
peak. 
%gb I am not sure of the use of peaking as a noun but left it - The peaking is the lowest 
The peaking is not as pronounced for a negative polarization since in
this case the decay products are mostly generated backwards in the
rest frame (cf. Eq.~\eqref{eq:topdecayeq}). At $\phi_l=\pi$ the order of peaking is inverted since we
are plotting normalized distributions. As expected, the benchmarks with a large mass
difference differentiate stronger between different polarizations than
small mass differences.
In general, the distributions in Fig.~\ref{fig:Phi_dist} seem to be
well separated by their polarization value. Therefore we
quantify this with an asymmetry parameter $A_\phi$ defined by
\begin{equation}
A_\phi=\frac{\sigma(\cos\phi_l>0)-\sigma(\cos\phi_l<0)}{\sigma(\cos\phi_l>0)
+\sigma(\cos\phi_l<0)}\,.
\label{Aldef}
\end{equation}
The polarization is influenced by the boost to the stop labframe (section \ref{sec:Stop_Boost}). We will treat the transverse momentum ($p_T$) of the top as a crude qualifier of this boost and apply a cut on $p_T$  \cite{Godbole:2010kr}. Thereby attempting to reduce the polluting effect of the kinematics on the angular distribution. We have defined an adaptive cut as
\begin{equation}
%\frac{p_{T,\mbox{max}}}{x} < p_T <  x\,p_{T\,\mbox{max}}\,.
\frac{p_T^{max}}{x} < p_T <  x\,p_T^{max}\,.
\label{eq:ptcut}
\end{equation}
We define both a strict ($x=1.5$) and loose ($x=2$) cut.
The results for these choices are given in Tab.~\ref{Aphi_bench2}.  \\
%gb Use a uniform notation for equations Eq.xx

\begin{table}[!h]
\centering
\begin{tabular}{c|ccc||ccc}
$P_t$&$A_\phi$ no cut&$A_\phi$ loose cut&$A_\phi$ strict cut&$A_\phi$ no cut&$A_\phi$ loose cut&$A_\phi$ strict cut\\
\hline
+1& 0.57 & 0.51 & 0.48    & 0.87 & 0.90 & 0.90  \\
+0.5& 0.53 & 0.45 & 0.41 & 0.81 & 0.84 & 0.84  \\
0& 0.48 & 0.42 & 0.39         & 0.69 & 0.67 & 0.64  \\
-0.5 & 0.44 & 0.37 & 0.34     & 0.61 & 0.60 & 0.58  \\
-1.0 & 0.39 & 0.33 & 0.29 & 0.55 & 0.50 & 0.46   
\end{tabular}
\caption{Relative azimuthal asymmetry parameter for the process as defined in Eq.~\ref{eq:stopantistop}.  The left side of the table denotes small mass differences and the right side large mass differences between stop and neutralino. 
% I would remove that because it is not exactly invariant and we said it already The differences are kept invariant for the different benchmarks while the polarization is varied. 
An adaptive cut is applied on the transverse momentum as defined in Eq.~\ref{eq:ptcut}.}\label{Aphi_bench2}
\end{table}

%\begin{table}[!h]
%\centering
%\begin{tabular}{c|ccc||ccc}
%$P_t$&$A_\phi^R$ no cut&$A_\phi^R$ loose cut&$A_\phi^R$ strict cut&$A_\phi^R$ no cut&$A_\phi^R$ loose cut&$A_\phi^R$ strict cut\\
%\hline
%+1& 8.83 & 9.41 & 9.39 & 18.1 & 23.3 & 26.2  \\
%+0.5& 4.21 & 2.74 & 2.58 & 12.6 & 17.2 & 19.7  \\
%0& 0.48 & 0.42 & 0.39     & 0.69 & 0.67 & 0.64  \\
%-0.5 & -4.89 & -4.76 & -5.03  & -7.63 & -6.59 & -6.45  \\
%-1.0 &  -8.98 & -9.36 & -9.45 & -13.9 & -16.6 & -18.2 
%\end{tabular}
%\end{table}

From Tab.~\ref{Aphi_bench2} we notice that the asymmetry parameter $A_\phi$ is large for positive
polarizations, decreases for lower polarizations and reaches
its lowest value at a negative polarization. As expected, the $p_T$ cut
improves the asymmetry parameter. In the case of a small mass
difference, the effect is small. For large mass differences however, the
two $p_T$ cuts enhance the separation of different polarizations. This is natural, as a large stop-neutralino mass difference
endows the top with more kinetic energy. 

\vspace{.5cm}
\noindent{\bf Polar  asymmetries}
\vspace{.2cm}

\begin{figure}[!h]
\centering
\includegraphics[angle=0,width=0.49\linewidth]{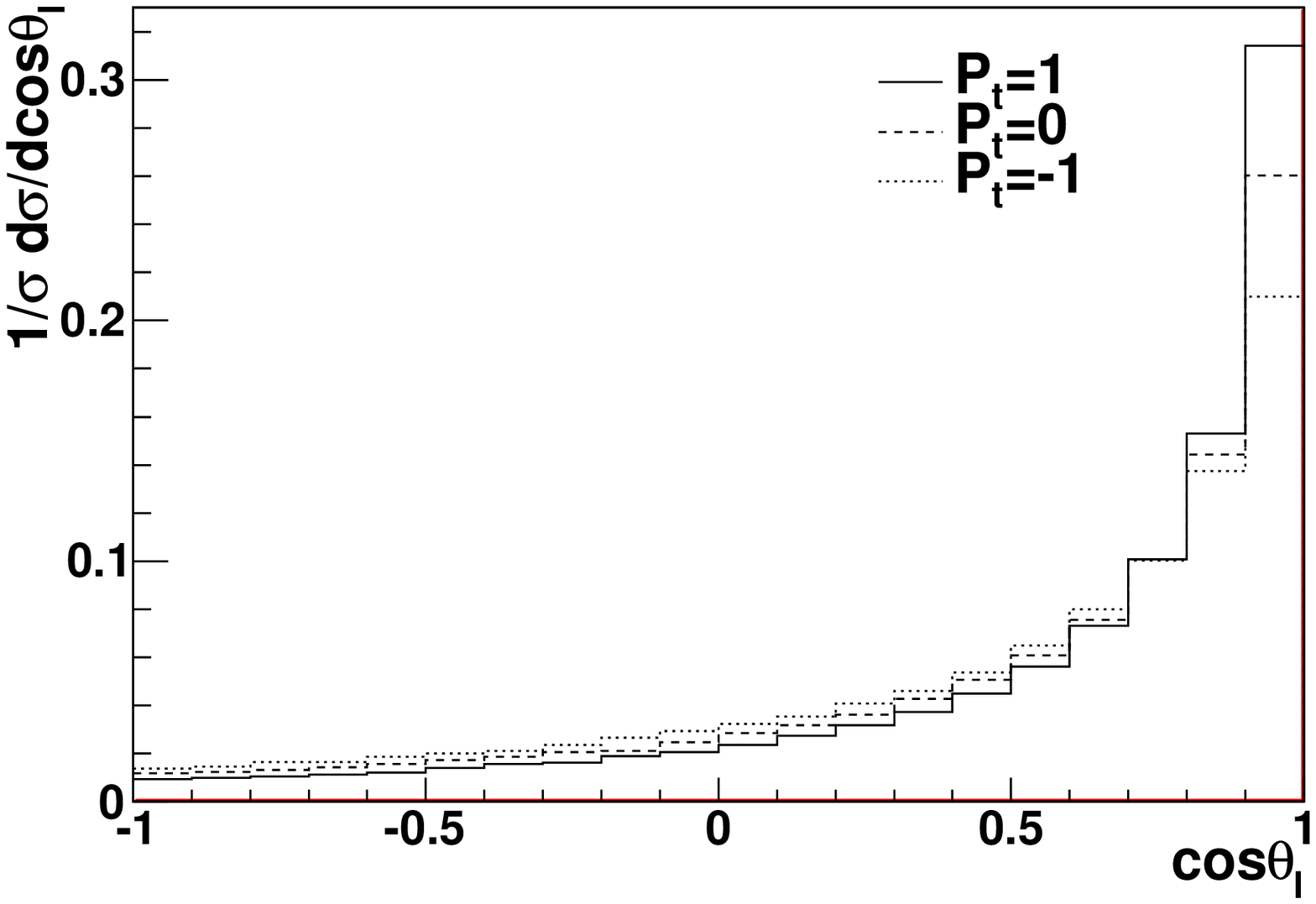}
\includegraphics[angle=0,width=0.49\linewidth]{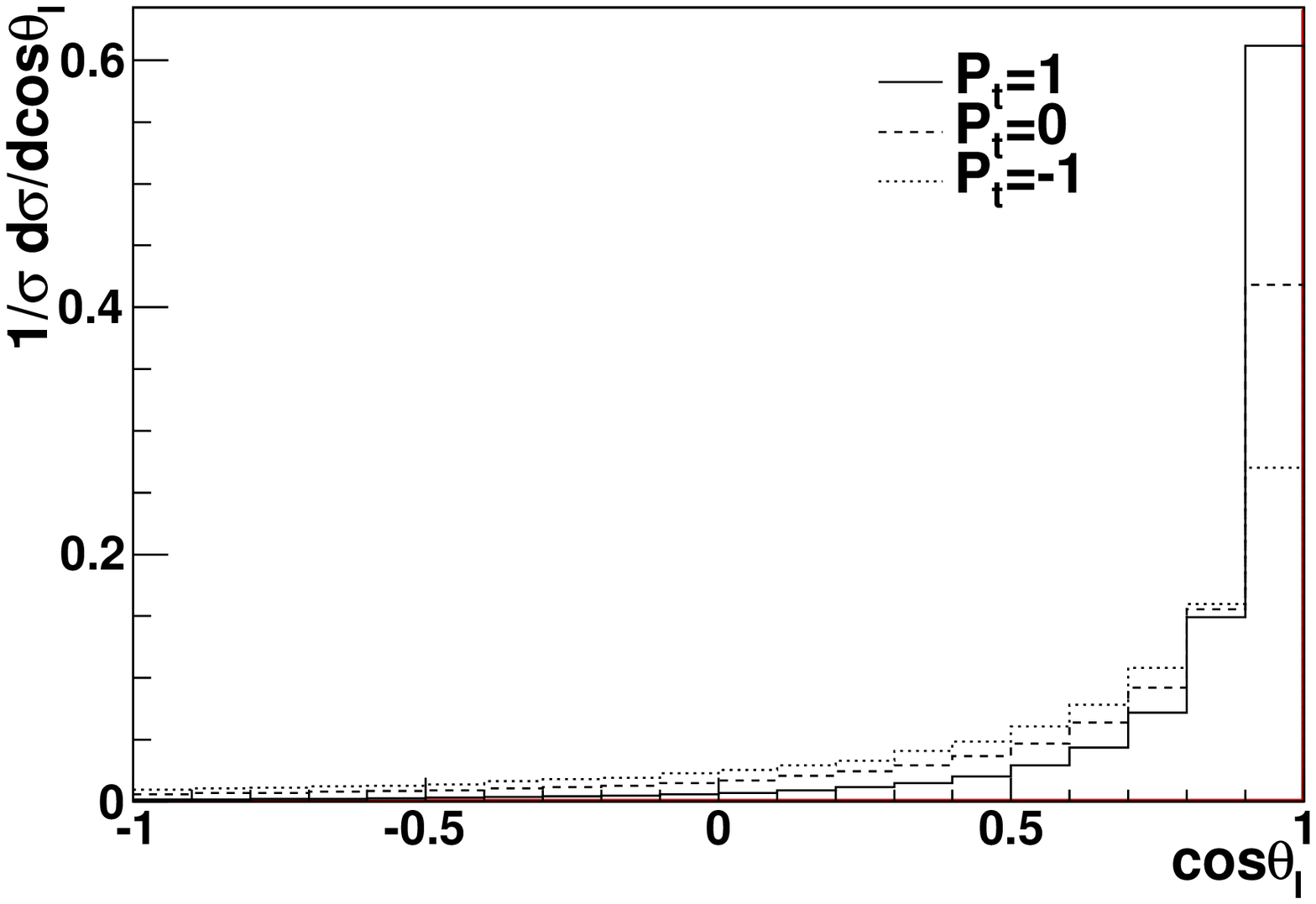}
\caption{The polar distribution $\theta_l$ of the decay lepton of the top quark. Polarizations in the left figure are chosen such that there is a small mass difference between stop and neutralino. In the right figure the mass difference is large. }\label{fig:Theta_dist}\end{figure}

We can apply a similar analysis to the distribution in the polar angle, defined as the angle between
top direction and decay lepton in the lab frame. The distributions are shown in 
Fig.~\ref{fig:Theta_dist}.  We notice a peaking in the direction of the top boost which is
again strongest for a positive polarization and weakest for a negative
polarization. Again the large mass difference cases 
%gb correlate stronger with the
show a stronger correlation with the
polarization $P_t$ than the small mass difference cases.
Because the distribution of $\theta_l$ is non-symmetric we have more choice for an asymmetry parameter definition
that quantifies the shape differences.  We have chosen
\begin{equation}
A_{\theta}=\frac{\sigma(\theta_l<\pi/4)-\sigma(\theta_l>\pi/4)}
{\sigma(\theta_l>\pi/4)+\sigma(\theta_l<\pi/4)}\,.
\label{Athetadef}
\end{equation}
The values for this parameter for various values of the  adaptive cut on $p_T$ are listed in Tab.~\ref{Atheta}.   

\begin{table}[!h]
\centering
\begin{tabular}{c|ccc||ccc}
$P_t$&$A_\theta$ no cut&$A_\theta$ loose cut&$A_\theta$ strict cut&$A_\theta$ no cut&$A_\theta$ loose cut&$A_\theta$ strict cut\\
\hline
+1& -0.12 & -0.02 & 0.02 & -0.66 & -0.66 & -0.64 \\ 
+0.5 & -0.06 & 0.05 & 0.08 &-0.55 & -0.55 & -0.52  \\
0& 0.001 & 0.10 & 0.13 & -0.32 & -0.25 & -0.20 \\ 
-0.5 & 0.06 & 0.14 & 0.17 &-0.18 & -0.12 & -0.07 \\
-1.0 & 0.12 & 0.20 & 0.22 & -0.06 & 0.03 & 0.10  
\end{tabular}
\caption{Relative polar asymmetry parameter for the process as defined in Eq.~\ref{eq:stopantistop}. The left side denotes benchmarks with a small mass difference and the right side large mass differences between stop and neutralino.  An adaptive cut is applied on the transverse momentum as defined in  Eq.~\ref{eq:ptcut}.}\label{Atheta}
\end{table}

%\begin{table}[!h]
%\centering
%\begin{tabular}{c|ccc||ccc}
%$P_t$&$A_\theta^R$ no cut&$A_\theta^R$ loose cut&$A_\theta^R$ strict cut&$A_\theta^R$ no cut&$A_\theta^R$ loose cut&$A_\theta^R$ strict cut\\
%\hline
%+1& -12.5 & -11.9 & -11.6 & -33.7 & -41.0 & -43.9  \\ 
%+0.5 & -6.28 & -4.49 & -5.04 & -23.4 & -30.3 & -32.7  \\
%0& 0.001 & 0.10 & 0.13 & -0.32 & -0.25 & -0.20 \\ 
%-0.5 & 6.16 & 4.84 & 4.16 &  14.1 & 12.5 & 12.3 \\
%-1.0 &  11.6 & 9.95 & 8.82 & 25.7 & 28.3 & 29.2   
%\end{tabular}
%\caption{Relative polar asymmetry parameter for the process as defined in Eq.~\ref{eq:stopantistop}. The left side denotes benchmarks with a small mass difference and the right side large mass differences between stop and neutralino.  An adaptive cut is applied on the transverse momentum as defined in  Eq.~\ref{eq:ptcut}.}\label{Atheta}
%\end{table}
We notice that $A_\theta$ may become negative. 
It is of course possible to define the asymmetry parameter such that
all values are positive. However, in an experimental analysis, the definition of $A_\theta$ will
be tuned to enhance the effects of polarization. As the outcome of
this procedure will depend on the masses of the sparticles, we will
use the definition given in \cite{Godbole:2011vw} to show the
qualitative effect.
The value of $A_\theta$ is lowest for positive
polarization, increases as the polarization decreases and reaches its
highest value at a polarization of $P_t=-1$. The adaptive cut again
has little effect for the small mass differences but enhances 
%gb
mildly the separation of $A_\theta$
for large mass differences and can therefore be a useful probe for polarization.  

%+1& -0.124085 & -0.022907 & 0.0156407 & -0.657203 & -0.655793 & -0.636253 \\ 
%+0.5 & -0.0613853 & 0.0511564 & 0.0815968 &-0.554899 & -0.548626 & -0.524415  \\
%0& 0.00140414 & 0.0960479 & 0.132016 & -0.320587 & -0.246121 & -0.197137 \\ 
%-0.5 & 0.0629766 & 0.144434 & 0.173569 &-0.179091 & -0.120735 & -0.0737752 \\
%-0.8 &  0.0943702 & 0.186129 & 0.218162 &                   &                     &                       \\
%-1.0 &                       &                   &                    & -0.0635179 & 0.032676 & 0.0953227  
\newpage
\vspace{0.5cm}
\noindent{\bf Impact of the stop neutralino mass difference}
\vspace{.2cm}

%gb Here I move the discussion on the mass difference because it will be relevant for the case of three neutralinos
We have seen in section \ref{sec:Masses} that the polarization depends
on the mass difference between the stop and the neutralino,  
more precisely on $\Delta m=m_{\tilde t_1}-m_{\tilde\chi^0_1}-m_t$,  and 
that the asymmetry parameter $A_\phi$ is highest for a high polarization for both mass differences. So far we have
been studying the effects of polarization whilst keeping this
difference constant.
%Table \ref{Aphi_bench1} shows the results for $A_\phi$ including the adaptive cut. 
However, when we vary $\Delta m$,  the asymmetry values corresponding to different polarizations are not well separated anymore. For example, we consider a new benchmark  with a mass difference that falls in between the two cases in Tab.~\ref{bench2}. For this point $P_t=0$, $m_{\tilde t}=508.9$ GeV, $m_{\tilde \chi^0_1}=292.4$ GeV, $\sin\theta_{\tilde t}=0.1234$, and yet the  asymmetry $A_\phi=0.56$ is quite similar to the value for 
the benchmark  $P_t=0.5$ in Tab.~\ref{Aphi_bench2} which has $A_\phi=0.53$.  
The mass difference for these points varies from $\Delta m =53$ GeV for the former and $\Delta m =6$ GeV for the latter.
% Imposing the adaptive cut on the $p_T$ of the top strengthens the signal, but neither cuts are able to isolate the purely %polarization induced behavior. We get  $A_\phi=0.55 (0.53)$ for the loose (strict) cut. We conclude that varying the mass %difference  slightly has a large effect on the angular distributions and therefore pollutes the information of polarization present %in these angular distributions. 
Imposing the adaptive cut on the $p_T$ of the top enhances the differences between the two benchmarks, but neither cuts are able to isolate the purely polarization induced behavior. For the $P_t=0$ benchmark we get  $A_\phi=0.55 (0.53)$ for the loose (strict) cut to be compared with $A_\phi=0.45(0.42)$ for the $P_t=0.5$ benchmark. We conclude that varying the mass difference  slightly has a large effect on the angular distributions and therefore pollutes the information of polarization present in these angular distributions. 
Thus, detailed mass measurements are needed in addition to the polarization-dependent observables.

\vspace{.5cm}
\noindent{\bf Decays to $\tilde\chi^0_2,\tilde\chi^0_3$}
\vspace{.2cm}

Thus far we have studied the case where the stop decays to one, generic neutralino type. We next
examine the case where we allow for a decay to multiple neutralino types. 
Two large mass difference benchmarks of Tab.~\ref{bench2} have stop branching ratios to several neutralino types, those  with A) $P_t=0.5$ and B) $P_t=0$.
% in Table \ref{bench1} both have decay modes of the stop into $t\tilde\chi^0_2, t\tilde\chi^0_3$. 
In  case A the heavier neutralino masses are $m_{\tilde \chi^0_2}=207$ GeV, $m_{\tilde \chi^0_3}=213$ GeV while in  case B, $m_{\tilde \chi^0_2}=276$ GeV, $m_{\tilde \chi^0_3}=282$ GeV.
The heavier neutralinos are higgsino-like so that the polarization is close to $P_t=-1$ in  case A which has a RH stop and to $P_t=1$ in case B with a LH stop. 
We have  listed the separate contributions to $P_t$ and the asymmetries  $A_\phi$ and $A_\theta$ in Tab.~\ref{Aphi_one_decay_Ptp5}. The difference in the asymmetries between various neutralino channels is somewhat less than naively expected. This is because the mass difference $\Delta m$ is smaller for heavier neutralinos, thus reducing the difference in the asymmetries as discussed above. This effect is particularly noticeable for the second case
where despite the fact that $P_t=0(1)$ for the light (heavier) neutralinos, all three neutralinos give rise to almost the same asymmetries. 

\begin{table}[!h]
\centering
\begin{tabular}{c|cccc|cccc|}
& &Case A&&&&Case B&&\\ 
 decay to & $A_\phi$ & $A_\theta$ & $P_t$ & BR & $A_\phi$ & $A_\theta$ & $P_t$ & BR \\
\hline
$\chi_1^0$ & 0.81 & -0.55  & 0.5 &6.5\%  &  0.69  & -0.31& 0.0 &2.7\% \\
$\chi_2^0$ & 0.53 & -0.04 & -1.0 & 20\%  & 0.71 &  -0.34& 0.99 & 29.3\%\\ 
$\chi_3^0$ & 0.53 &  -0.05 & -0.88 & 18\%&  0.69 & -0.31& 0.96& 29.8\%
\end{tabular}
\caption{Azimuthal and polar asymmetry parameter for the process as defined in Eq.~(\ref{eq:stopantistop}) allowing for decays of the stop to a certain neutralino type. The polarization and branching fraction for the decay into each neutralino channel is also specified. Case A and Case B 
correspond respectively to the second and third  rows of the large mass difference benchmarks in Tab.~\ref{bench2}  .}
\label{Aphi_one_decay_Ptp5}
\end{table}

 With the theoretical prediction on the rest frame polarization per decay mode on the basis of Eqs.~\ref{eq:fs} and \ref{eq:stopcouplings}, the lab frame distributions can then be predicted after  combination with the appropriate Lorentz transformations. The asymmetry parameter for all decays is a sum of the individual values weighted by branching ratios. 
 The extent to which $A_\phi$ depends on the angular distribution of a certain decay mode therefore depends strongly on the branching fractions.
 The results including adaptive cuts for the two benchmarks of Tab.~\ref{Aphi_one_decay_Ptp5} are shown in Tab.~\ref{Aphi_all_decay}.    Clearly the asymmetries are dominated by the heavier neutralino decay channels for case A while they receive similar contributions from all three neutralino channels for  case B.

\begin{table}[!h]
\centering
\begin{tabular}{c|c|ccc||ccc}
&$P_t(t\tilde\chi^0_1)$&$A_\phi$ no cut&$A_\phi$ loose cut&$A_\phi$ strict cut&$A_\theta$ no cut&$A_\theta$ loose cut&$A_\theta$ strict cut\\
\hline
A&+0.5 & 0.58 & 0.53 & 0.50 & -0.13 & -0.03 & 0.02  \\
B& 0       & 0.70 & 0.69 & 0.68 &   -0.32 & -0.26 & -0.22
\end{tabular}
\caption{Azimuthal and polar asymmetry parameter for the process as defined in Eq.~(\ref{eq:stopantistop}( allowing for decays of the stop to all neutralino types.  An adaptive cut is applied on the transverse momentum as defined in Eq.~(\ref{eq:ptcut}).}
\label{Aphi_all_decay}
\end{table}

\section{Conclusion}\label{s:conclusion}

The phenomenology of the third generation sfermions  has always been an interesting subject to explore as
this can yield non-trivial information about SUSY parameters. 
In view of the ever increasing upper limits on the masses of
the strongly interacting sparticles that are being extracted from LHC data and the observation of a light,
single Higgs-like particle  naturalness considerations within the MSSM leads to the possibility 
of  third  generation sfermions that are much lighter than the first two generations. 
Thus  direct pair production cross-sections of both stops and sbottoms can be large enough to be probed within the current run of the LHC. The top quarks produced
in these decays are generally polarized and this polarization  holds
information about mixing in the squark  sector,  mixing in the
chargino/neutralino sectors as well as on the top velocity, hence on the mass difference between the squark and the neutralino/chargino.  
The parameters that affect the top polarization will influence the effectiveness of the searches for stops. Thus, the limits extracted will not only depend on the stop and neutralino mass but also on the assumed polarization. Indeed, 
the polarization can affect the energies of decay leptons and hence the optimization of cuts
to reduce the background from the QCD produced unpolarized top. 
Since the top polarization goes to zero in the limit of a small
stop-neutralino mass difference, the polarization-induced kinematic
effects will be particularly important for models where this mass
difference is large. This is an important factor to keep in mind in
analyses using simplified models with large mass differences. To
obtain a conservative limit, one should use a model which produces a
completely negatively polarized top quark.

We have explored the
possible values of the top polarization in the decay of the lightest stop into a top and a neutralino  and we have scanned the parameter space which is consistent with a light Higgs.
We find that the bino content of the neutralino is a critical parameter  and that due to the largeness of the hypercharge for the right-handed top which  drives the  bino-stop-top coupling,  a mixed stop often behaves like a RH stop. 
 A dominantly RH stop produces a negative top polarization when it decays into a higgsino and a positive polarization when the decay is into a bino, and vice-versa for a LH stop.
 This implies that positive top polarization leads to more energetic leptons,  allowing for events to be separated more easily from the top pair background. The LH stop with a higgsino LSP and the RH stop with a bino LSP could be more tightly constrained at the LHC than the other two combinations. We have also shown that although small branching ratios into the lightest neutralino can occur especially for the decay into a higgsino, similar polarizations for the decay into the two higgsino states imply that we can exploit both decay modes to measure the top polarization. 
Finally, a small  mass difference between the stop and the neutralino leads to a very small polarization.

We analyzed the kinematics of the decay products of the  top arising from stop decay into a top and a neutralino in the laboratory frame. Since the majority of the top quarks in the SM background are unpolarized the stop search is particularly challenged in the $t \tilde \chi_{1}^{0}$ mode for points
in the parameter space which give rise to tops with negative polarization. The spectrum of the electron energy as well as transverse momentum of the lepton, softens (hardens) for negatively (positively) polarized top quarks respectively, compared to an unpolarized top quark. This modification of the
position of the peak increases with increasing value of 
$m_{\tilde t} - m_{\tilde \chi_{1}^{0}}$. For the electron energy spectrum the shift is $-30$ GeV for 
$m_{\tilde t} - m_{\tilde \chi_{1}^{0}} \sim 320$ GeV and 
$-16$  GeV for  $m_{\tilde t} - m_{\tilde \chi_{1}^{0}} \sim 130$ GeV. Thus we see that even with the same kinematics, the 
reach of a particular search using the lepton is less efficient for
negatively polarized tops. This effect is more pronounced for large mass differences between the stop and the neutralino.
 
Finally, we have studied lab-frame observables and
defined  asymmetries in the polar and azimuthal angle. These asymmetries have both a polarization-dependent and independent part and provide a useful probe for top polarization provided the masses of the particles involved are known, since the polarization is very sensitive to mass differences.
In conclusion, study of the top polarization can provide useful information on supersymmetric parameters at the LHC when the supersymmetric partner of the top is discovered.

\section*{Acknowledgments}
We thank Wim Beenakker and Eric Laenen for many useful discussions. IN and LH are supported by the Foundation for Fundamental Research of Matter (FOM), program 104 ``Theoretical Particle Physics in the Era of the LHC''. RG wishes to acknowledge the
Department of Science and Technology of India,
for financial support  under the J.C. Bose Fellowship 
scheme under grant no. SR/S2/JCB-64/2007
and NIKHEF as well as LPSC, Grenoble for 
hospitality when part of this work was done.

\noindent
{\bf Note added:} As this paper was finalised new results from direct stop searches were presented by ATLAS including $13fb^{-1}$ of data from the 8TeV run~\cite{LHCC_stop}. These extend the stop exclusion to 580 GeV for massless neutralinos. 
When they decay exclusively into $t \tilde\chi^0_1$ stops of 500 GeV are excluded if  the neutralino LSP is lighter than 200 GeV.
These limits can be somewhat weakened because of smaller branching ratios as well as because of top polarization effects. 
Only one of the benchmark we have used in this paper falls within the ATLAS exclusion, the one with $P_t=1$ and large mass difference.

\bibliography{refs-r}

\providecommand{\href}[2]{#2}\begingroup\raggedright\begin{thebibliography}{10}

\bibitem{ATLAS:2012gk}
{\bf ATLAS} Collaboration, G.~Aad {\em et al.}, ``{Observation of a new
  particle in the search for the Standard Model Higgs boson with the ATLAS
  detector at the LHC},'' {\em Phys.Lett.B} (2012)
\href{http://www.arXiv.org/abs/1207.7214}{{\tt 1207.7214}}.
%%CITATION = ARXIV:1207.7214;%%.

\bibitem{CMS:2012gu}
{\bf CMS} Collaboration, S.~Chatrchyan {\em et al.}, ``{Observation of a new
  boson at a mass of 125 GeV with the CMS experiment at the LHC},'' {\em
  Phys.Lett.B} (2012)
\href{http://www.arXiv.org/abs/1207.7235}{{\tt 1207.7235}}.
%%CITATION = ARXIV:1207.7235;%%.

\bibitem{Wess:1974tw}
J.~Wess and B.~Zumino, ``{Supergauge Transformations in Four-Dimensions},''
  {\em Nucl.Phys.} {\bf B70} (1974) 39--50.

\bibitem{Nilles:1983ge}
H.~P. Nilles, ``{Supersymmetry, Supergravity and Particle Physics},'' {\em
  Phys. Rept.} {\bf 110} (1984)
1.
%%CITATION = PRPLC,110,1;%%.

\bibitem{Aad:2011ib}
{\bf ATLAS} Collaboration, G.~Aad {\em et al.}, ``{Search for Squarks and
  Gluinos Using Final States with Jets and Missing Transverse Momentum with the
  ATLAS Detector in $\sqrt{s}$ = 7 TeV Proton-Proton Collisions},'' {\em Phys.
  Lett.} {\bf B710} (2012) 67,
\href{http://www.arXiv.org/abs/1109.6572}{{\tt 1109.6572}}.
%%CITATION = ARXIV:1109.6572;%%.

\bibitem{ATLAS-CONF-2012-109}
{\bf ATLAS} Collaboration, ``{Search for squarks and gluinos with the ATLAS
  detector using final states with jets and missing transverse momentum and
  $5.8 fb^{-1}$ of $\sqrt{s}=8 {\rm TeV}$ proton-proton collision data},''
  2012.
\newblock { ATLAS-CONF-2012-133, see
  \texttt{http://atlas.web.cern.ch/Atlas/GROUPS/PHYSICS/CONFNOTES/}}.

\bibitem{Chatrchyan:2011zy}
{\bf CMS} Collaboration, S.~Chatrchyan {\em et al.}, ``{Search for
  Supersymmetry at the LHC in Events with Jets and Missing Transverse
  Energy},'' {\em Phys. Rev. Lett.} {\bf 107} (2011) 221804,
\href{http://www.arXiv.org/abs/1109.2352}{{\tt 1109.2352}}.
%%CITATION = ARXIV:1109.2352;%%.

\bibitem{Chatrchyan:2012jx}
{\bf CMS} Collaboration, S.~Chatrchyan {\em et al.}, ``{Search for
  supersymmetry in hadronic final states using MT2 in $pp$ collisions at
  $\sqrt{s} = 7$ TeV},'' {\em JHEP} {\bf 1210} (2012) 018,
\href{http://www.arXiv.org/abs/1207.1798}{{\tt 1207.1798}}.
%%CITATION = ARXIV:1207.1798;%%.

\bibitem{Fischler:1981zk}
W.~Fischler, H.~P. Nilles, J.~Polchinski, S.~Raby, and L.~Susskind,
  ``{Vanishing Renormalization of the D Term in Supersymmetric U(1)
  Theories},'' {\em Phys.Rev.Lett.} {\bf 47} (1981)
757.
%%CITATION = PRLTA,47,757;%%.

\bibitem{Kaul:1981tp}
R.~K. Kaul and P.~Majumdar, ``{Naturalness in a globally supersymmetric gauge
  theory with elementary scalar fields},'' 1981.
\newblock {Print-81-0373 (BANGALORE)}.

\bibitem{Kaul:1981hi}
R.~K. Kaul and P.~Majumdar, ``{Cancellation of Quadratically divergent mass
  corrections in globally supersymmetric spontaneously broken gauge
  theories},'' {\em Nucl.Phys.} {\bf B199} (1982)
36.
%%CITATION = NUPHA,B199,36;%%.

\bibitem{Kaul:1981wp}
R.~K. Kaul, ``{Gauge Hierarchy in a Supersymmetric Model},'' {\em Phys.Lett.}
  {\bf B109} (1982)
19.
%%CITATION = PHLTA,B109,19;%%.

\bibitem{Dimopoulos:1981zb}
S.~Dimopoulos and H.~Georgi, ``{Softly Broken Supersymmetry and SU(5)},'' {\em
  Nucl.Phys.} {\bf B193} (1981)
150.
%%CITATION = NUPHA,B193,150;%%.

\bibitem{Barbieri:1987fn}
R.~Barbieri and G.~Giudice, ``{Upper Bounds on Supersymmetric Particle
  Masses},'' {\em Nucl.Phys.} {\bf B306} (1988)
63.
%%CITATION = NUPHA,B306,63;%%.

\bibitem{deCarlos:1993yy}
B.~de~Carlos and J.~Casas, ``{One loop analysis of the electroweak breaking in
  supersymmetric models and the fine tuning problem},'' {\em Phys.Lett.} {\bf
  B309} (1993) 320--328,
\href{http://www.arXiv.org/abs/hep-ph/9303291}{{\tt hep-ph/9303291}}.
%%CITATION = HEP-PH/9303291;%%.

\bibitem{Brust:2011tb}
C.~Brust, A.~Katz, S.~Lawrence, and R.~Sundrum, ``{SUSY, the Third Generation
  and the LHC},'' {\em JHEP} {\bf 1203} (2012) 103,
\href{http://www.arXiv.org/abs/1110.6670}{{\tt 1110.6670}}.
%%CITATION = ARXIV:1110.6670;%%.

\bibitem{Arbey:2012dq}
A.~Arbey, M.~Battaglia, A.~Djouadi, and F.~Mahmoudi, ``{The Higgs sector of the
  phenomenological MSSM in the light of the Higgs boson discovery},'' {\em
  JHEP} {\bf 1209} (2012) 107,
\href{http://www.arXiv.org/abs/1207.1348}{{\tt 1207.1348}}.
%%CITATION = ARXIV:1207.1348;%%.

\bibitem{Barger:2012hr}
V.~Barger, P.~Huang, M.~Ishida, and W.-Y. Keung, ``{Scalar-Top Masses from SUSY
  Loops with 125 GeV mh and Precise Mw},''
\href{http://www.arXiv.org/abs/1206.1777}{{\tt 1206.1777}}.
%%CITATION = ARXIV:1206.1777;%%.

\bibitem{Beenakker:1996ed}
W.~Beenakker, R.~Hopker, and M.~Spira, ``{PROSPINO: A Program for the
  production of supersymmetric particles in next-to-leading order QCD},''
\href{http://www.arXiv.org/abs/hep-ph/9611232}{{\tt hep-ph/9611232}}.
%%CITATION = HEP-PH/9611232;%%.

\bibitem{Beenakker:2010nq}
W.~Beenakker, S.~Brensing, M.~Kr\"amer, A.~Kulesza, E.~Laenen, and I.~Niessen,
  ``{Supersymmetric Top and Bottom Squark Production at Hadron Colliders},''
  {\em JHEP} {\bf 8} (2010) 1,
\href{http://www.arXiv.org/abs/1006.4771}{{\tt 1006.4771}}.
%%CITATION = 1006.4771;%%.

\bibitem{Beenakker:2011fu}
W.~Beenakker, S.~Brensing, M.~Kr\"amer, A.~Kulesza, E.~Laenen, L.~Motyka, and
  I.~Niessen, ``{Squark and Gluino Hadroproduction},'' {\em Int. J. Mod. Phys.}
  {\bf A26} (2011) 2637,
\href{http://www.arXiv.org/abs/1105.1110}{{\tt 1105.1110}}.
%%CITATION = 1105.1110;%%.

\bibitem{NLLfast}
W.~Beenakker, S.~Brensing, M.~Kr\"amer, A.~Kulesza, E.~Laenen, and I.~Niessen,
  ``{\texttt{NLL-fast}, a computer program which computes the squark and gluino
  hadroproduction cross sections including NLO SUSY-QCD corrections and the
  resummation of soft gluon emission at NLL accuracy}.'' See
  \texttt{http://web.physik}
  \texttt{.rwth-aachen.de/service/wiki/bin/view/kraemer/squarksandgluinos}.

\bibitem{sparticles}
M.~Drees, R.~Godbole, and P.~Roy, {\em Theory and Phenomenology of Sparticles}.
\newblock World Scientific, 2004.

\bibitem{Desai:2011th}
N.~Desai and B.~Mukhopadhyaya, ``{Constraints on supersymmetry with light third
  family from LHC data},'' {\em JHEP} {\bf 1205} (2012) 057,
\href{http://www.arXiv.org/abs/1111.2830}{{\tt 1111.2830}}.
%%CITATION = ARXIV:1111.2830;%%.

\bibitem{He:2011tp}
B.~He, T.~Li, and Q.~Shafi, ``{Impact of LHC Searches on NLSP Top Squark and
  Gluino Mass},'' {\em JHEP} {\bf 1205} (2012) 148,
\href{http://www.arXiv.org/abs/1112.4461}{{\tt 1112.4461}}.
%%CITATION = ARXIV:1112.4461;%%.

\bibitem{Drees:2012dd}
M.~Drees, M.~Hanussek, and J.~S. Kim, ``{Light Stop Searches at the LHC with
  Monojet Events},'' {\em Phys.Rev.} {\bf D86} (2012) 035024,
\href{http://www.arXiv.org/abs/1201.5714}{{\tt 1201.5714}}.
%%CITATION = ARXIV:1201.5714;%%.

\bibitem{Berger:2012ec}
J.~Berger, J.~Hubisz, and M.~Perelstein, ``{A Fermionic Top Partner:
  Naturalness and the LHC},'' {\em JHEP} {\bf 1207} (2012) 016,
\href{http://www.arXiv.org/abs/1205.0013}{{\tt 1205.0013}}.
%%CITATION = ARXIV:1205.0013;%%.

\bibitem{Plehn:2012pr}
T.~Plehn, M.~Spannowsky, and M.~Takeuchi, ``{Stop searches in 2012},'' {\em
  JHEP} {\bf 1208} (2012) 091,
\href{http://www.arXiv.org/abs/1205.2696}{{\tt 1205.2696}}.
%%CITATION = ARXIV:1205.2696;%%.

\bibitem{Han:2012fw}
Z.~Han, A.~Katz, D.~Krohn, and M.~Reece, ``{(Light) Stop Signs},'' {\em JHEP}
  {\bf 1208} (2012) 083,
\href{http://www.arXiv.org/abs/1205.5808}{{\tt 1205.5808}}.
%%CITATION = ARXIV:1205.5808;%%.

\bibitem{Berger:2012an}
E.~L. Berger, Q.-H. Cao, J.-H. Yu, and H.~Zhang, ``{Measuring Top Quark
  Polarization in Top Pair plus Missing Energy Events},''
\href{http://www.arXiv.org/abs/1207.1101}{{\tt 1207.1101}}.
%%CITATION = ARXIV:1207.1101;%%.

\bibitem{Bhattacherjee:2012ir}
B.~Bhattacherjee, S.~K. Mandal, and M.~Nojiri, ``{Top Polarization and Stop
  Mixing from Boosted Jet Substructure},''
\href{http://www.arXiv.org/abs/1211.7261}{{\tt 1211.7261}}.
%%CITATION = ARXIV:1211.7261;%%.

\bibitem{stop-atlas-prl-1:2012si}
{\bf ATLAS} Collaboration, G.~Aad {\em et al.}, ``{Search for a supersymmetric
  partner to the top quark in final states with jets and missing transverse
  momentum at $\sqrt{s}=7$ TeV with the ATLAS detector},''
\href{http://www.arXiv.org/abs/1208.1447}{{\tt 1208.1447}}.
%%CITATION = ARXIV:1208.1447;%%.

\bibitem{stop-atlas-prl-2:2012ar}
{\bf ATLAS} Collaboration, G.~Aad {\em et al.}, ``{Search for direct top squark
  pair production in final states with one isolated lepton, jets, and missing
  transverse momentum in $\sqrt{s}=7$ TeV $pp$ collisions using 4.7 $fb^{-1}$
  of ATLAS data},''
\href{http://www.arXiv.org/abs/1208.2590}{{\tt 1208.2590}}.
%%CITATION = ARXIV:1208.2590;%%.

\bibitem{Aad:2012uu}
{\bf ATLAS} Collaboration, G.~Aad {\em et al.}, ``{Search for a heavy top-quark
  partner in final states with two leptons with the ATLAS detector at the
  LHC},'' {\em JHEP} {\bf 1211} (2012) 094,
\href{http://www.arXiv.org/abs/1209.4186}{{\tt 1209.4186}}.
%%CITATION = ARXIV:1209.4186;%%.

\bibitem{Aad:2012yr}
{\bf ATLAS} Collaboration, G.~Aad {\em et al.}, ``{Search for light top squark
  pair production in final states with leptons and $b^-$ jets with the ATLAS
  detector in $\sqrt{s}=7$ TeV proton-proton collisions},''
\href{http://www.arXiv.org/abs/1209.2102}{{\tt 1209.2102}}.
%%CITATION = ARXIV:1209.2102;%%.

\bibitem{CMS-PAS-SUS-12-023}
{\bf CMS} Collaboration, S.~Chatrchyan {\em et al.}, ``Search for direct top
  squark pair production in events with a single isolated lepton, jets and
  missing transverse energy at $\sqrt s = 8$ tev,'' 2012.
\newblock {CMS-PAS-SUS-12-023}.

\bibitem{CMS-PAS-SUS-11-030}
{\bf CMS} Collaboration, S.~Chatrchyan {\em et al.}, ``Scalar top quark search
  with jets and missing momentum in pp collisions at $\sqrt s = 7$ tev,'' 2012.
\newblock {CMS-PAS-SUS-11-030}.

\bibitem{Chatrchyan:2012sv}
{\bf CMS} Collaboration, S.~Chatrchyan {\em et al.}, ``{Search for pair
  production of third-generation leptoquarks and top squarks in $pp$ collisions
  at $\sqrt{s}=7$ TeV},''
\href{http://www.arXiv.org/abs/1210.5629}{{\tt 1210.5629}}.
%%CITATION = ARXIV:1210.5629;%%.

\bibitem{Weber:1479435}
H.~A. Weber, ``Search for third generation squarks at the lhc (status of light
  stop and sbottom searches),'' Tech. Rep. CMS-CR-2012-184.
  CERN-CMS-CR-2012-184, CERN, Geneva, Aug, 2012.

\bibitem{Nojiri:1994it}
M.~M. Nojiri, ``{Polarization of $\tau$ lepton from scalar $\tau$ decay as a
  probe of neutralino mixing},'' {\em Phys.Rev.} {\bf D51} (1995) 6281--6291,
\href{http://www.arXiv.org/abs/hep-ph/9412374}{{\tt hep-ph/9412374}}.
%%CITATION = HEP-PH/9412374;%%.

\bibitem{Godbole:2011vw}
R.~M. Godbole, L.~Hartgring, I.~Niessen, and C.~D. White, ``{Top Polarisation
  Studies in $H^-t$ and $Wt$ Production},'' {\em JHEP} {\bf 1} (2012) 011,
\href{http://www.arXiv.org/abs/1111.0759}{{\tt 1111.0759}}.
%%CITATION = 1111.0759;%%.

\bibitem{Hikasa:1999wy}
K.-i. Hikasa, J.~M. Yang, and B.-L. Young, ``{R-parity violation and top quark
  polarization at the Fermilab Tevatron collider},'' {\em Phys.Rev.} {\bf D60}
  (1999) 114041, \href{http://www.arXiv.org/abs/hep-ph/9908231}{{\tt
  hep-ph/9908231}}.

\bibitem{Li:2006he}
P.-Y. Li, G.-R. Lu, J.~M. Yang, and H.~Zhang, ``{Probing R-parity Violating
  Interactions from Top Quark Polarization at LHC},'' {\em Eur.Phys.J.} {\bf
  C51} (2007) 163--168, \href{http://www.arXiv.org/abs/hep-ph/0608223}{{\tt
  hep-ph/0608223}}.

\bibitem{Arai:2010ci}
M.~Arai, K.~Huitu, S.~K. Rai, and K.~Rao, ``{Single production of sleptons with
  polarized tops at the Large Hadron Collider},'' {\em JHEP} {\bf 1008} (2010)
  082, \href{http://www.arXiv.org/abs/1003.4708}{{\tt 1003.4708}}.

\bibitem{Cao:2010nw}
J.~Cao, L.~Wu, and J.~M. Yang, ``{New physics effects on top quark spin
  correlation and polarization at the LHC: a comparative study in different
  models},'' {\em Phys.Rev.} {\bf D83} (2011) 034024,
  \href{http://www.arXiv.org/abs/1011.5564}{{\tt 1011.5564}}.

\bibitem{Jung:2010yn}
D.-W. Jung, P.~Ko, and J.~S. Lee, ``{Longitudinal top polarization as a probe
  of a possible origin of forward-backward asymmetry of the top quark at the
  Tevatron},'' {\em Phys.Lett.} {\bf B701} (2011) 248--254,
  \href{http://www.arXiv.org/abs/1011.5976}{{\tt 1011.5976}}.

\bibitem{Choudhury:2010cd}
D.~Choudhury, R.~M. Godbole, S.~D. Rindani, and P.~Saha, ``{Top polarization,
  forward-backward asymmetry and new physics},'' {\em Phys.Rev.} {\bf D84}
  (2011) 014023, \href{http://www.arXiv.org/abs/1012.4750}{{\tt 1012.4750}}.

\bibitem{Krohn:2011tw}
D.~Krohn, T.~Liu, J.~Shelton, and L.-T. Wang, ``{A Polarized View of the Top
  Asymmetry},'' \href{http://www.arXiv.org/abs/1105.3743}{{\tt 1105.3743}}.

\bibitem{Boos:2003vf}
E.~Boos, H.~Martyn, G.~A. Moortgat-Pick, M.~Sachwitz, A.~Sherstnev, {\em et
  al.}, ``{Polarization in sfermion decays: Determining tan beta and trilinear
  couplings},'' {\em Eur.Phys.J.} {\bf C30} (2003) 395--407,
  \href{http://www.arXiv.org/abs/hep-ph/0303110}{{\tt hep-ph/0303110}}.

\bibitem{Gajdosik:2004ed}
T.~Gajdosik, R.~M. Godbole, and S.~Kraml, ``{Fermion polarization in sfermion
  decays as a probe of CP phases in the MSSM},'' {\em JHEP} {\bf 0409} (2004)
  051, \href{http://www.arXiv.org/abs/hep-ph/0405167}{{\tt hep-ph/0405167}}.

\bibitem{Shelton:2008nq}
J.~Shelton, ``{Polarized tops from new physics: signals and observables},''
  {\em Phys. Rev.} {\bf D79} (2009) 014032,
\href{http://www.arXiv.org/abs/0811.0569}{{\tt 0811.0569}}.
%%CITATION = 0811.0569;%%.

\bibitem{Perelstein:2008zt}
M.~Perelstein and A.~Weiler, ``{Polarized Tops from Stop Decays at the LHC},''
  {\em JHEP} {\bf 0903} (2009) 141,
  \href{http://www.arXiv.org/abs/0811.1024}{{\tt 0811.1024}}.

\bibitem{Krohn:2009wm}
D.~Krohn, J.~Shelton, and L.-T. Wang, ``{Measuring the Polarization of Boosted
  Hadronic Tops},'' {\em JHEP} {\bf 07} (2010) 041,
\href{http://www.arXiv.org/abs/0909.3855}{{\tt 0909.3855}}.
%%CITATION = 0909.3855;%%.

\bibitem{ATLASCONF-133}
{\bf ATLAS} Collaboration, ``{Measurement of top polarisation in $t \bar t$
  events},'' 2012.
\newblock {ATLAS-CONF-2012-133}.

\bibitem{Jezabek:1988ja}
M.~Jezabek and J.~H. Kuhn, ``{Lepton Spectra from Heavy Quark Decay},'' {\em
  Nucl.Phys.} {\bf B320} (1989) 20.

\bibitem{Czarnecki:1990pe}
A.~Czarnecki, M.~Jezabek, and J.~H. Kuhn, ``{Lepton spectra from decays of
  polarized top quarks},'' {\em Nucl.Phys.} {\bf B351} (1991) 70--80.

\bibitem{Brandenburg:2002xr}
A.~Brandenburg, Z.~Si, and P.~Uwer, ``{QCD corrected spin analyzing power of
  jets in decays of polarized top quarks},'' {\em Phys.Lett.} {\bf B539} (2002)
  235--241, \href{http://www.arXiv.org/abs/hep-ph/0205023}{{\tt
  hep-ph/0205023}}.

\bibitem{Grzadkowski:1999iq}
B.~Grzadkowski and Z.~Hioki, ``{New hints for testing anomalous top quark
  interactions at future linear colliders},'' {\em Phys.Lett.} {\bf B476}
  (2000) 87--94, \href{http://www.arXiv.org/abs/hep-ph/9911505}{{\tt
  hep-ph/9911505}}.

\bibitem{Grzadkowski:2002gt}
B.~Grzadkowski and Z.~Hioki, ``{Decoupling of anomalous top decay vertices in
  angular distribution of secondary particles},'' {\em Phys.Lett.} {\bf B557}
  (2003) 55--59, \href{http://www.arXiv.org/abs/hep-ph/0208079}{{\tt
  hep-ph/0208079}}.

\bibitem{Grzadkowski:2001tq}
B.~Grzadkowski and Z.~Hioki, ``{Angular distribution of leptons in general $t
  \bar{t}$ production and decay},'' {\em Phys.Lett.} {\bf B529} (2002) 82--86,
  \href{http://www.arXiv.org/abs/hep-ph/0112361}{{\tt hep-ph/0112361}}.

\bibitem{Hioki:2002vg}
Z.~Hioki, ``{A New decoupling theorem in top quark physics},'' in {\em
  {Seogwipo 2002, Linear colliders}}, pp.~333--338.
\newblock 2002.
\newblock \href{http://www.arXiv.org/abs/hep-ph/0210224}{{\tt hep-ph/0210224}}.

\bibitem{Ohkuma:2002iv}
K.~Ohkuma, ``{Effects of top quark anomalous decay couplings at gamma gamma
  colliders},'' {\em Nucl.Phys.Proc.Suppl.} {\bf 111} (2002) 285--287,
  \href{http://www.arXiv.org/abs/hep-ph/0202126}{{\tt hep-ph/0202126}}.

\bibitem{Rindani:2000jg}
S.~D. Rindani, ``{Effect of anomalous t b W vertex on decay lepton
  distributions in e+ e- $\to$ t anti-t and CP violating asymmetries},'' {\em
  Pramana} {\bf 54} (2000) 791--812,
  \href{http://www.arXiv.org/abs/hep-ph/0002006}{{\tt hep-ph/0002006}}.

\bibitem{Godbole:2002qu}
R.~M. Godbole, S.~D. Rindani, and R.~K. Singh, ``{Study of CP property of the
  Higgs at a photon collider using gamma gamma $\to$ t anti-t $\to$ lX},'' {\em
  Phys.Rev.} {\bf D67} (2003) 095009,
  \href{http://www.arXiv.org/abs/hep-ph/0211136}{{\tt hep-ph/0211136}}.

\bibitem{Godbole:2006tq}
R.~M. Godbole, S.~D. Rindani, and R.~K. Singh, ``{Lepton distribution as a
  probe of new physics in production and decay of the t quark and its
  polarization},'' {\em JHEP} {\bf 12} (2006) 021,
\href{http://www.arXiv.org/abs/hep-ph/0605100}{{\tt hep-ph/0605100}}.
%%CITATION = HEP-PH/0605100;%%.

\bibitem{Godbole:2010kr}
R.~M. Godbole, K.~Rao, S.~D. Rindani, and R.~K. Singh, ``{On measurement of top
  polarization as a probe of $t \bar t$ production mechanisms at the LHC},''
  {\em JHEP} {\bf 11} (2010) 144,
\href{http://www.arXiv.org/abs/1010.1458}{{\tt 1010.1458}}.
%%CITATION = 1010.1458;%%.

\bibitem{Godbole:2009dp}
R.~M. Godbole, S.~D. Rindani, K.~Rao, and R.~K. Singh, ``{Top polarization as a
  probe of new physics},'' {\em AIP Conf.Proc.} {\bf 1200} (2010) 682--685,
  \href{http://www.arXiv.org/abs/0911.3622}{{\tt 0911.3622}}.

\bibitem{Bouchiat1958416}
C.~Bouchiat and L.~Michel, ``Mesure de la polarisation des electrons
  relativistes,'' {\em Nuclear Physics} {\bf 5} (1958) 416 -- 434.

\bibitem{springerlink:10.1007/BF03026451}
L.~Michel, ``Covariant description of polarization,'' {\em Il Nuovo Cimento
  (1955-1965)} {\bf 14} (1959) 95--104.

\bibitem{Alwall:2011uj}
J.~Alwall, M.~Herquet, F.~Maltoni, O.~Mattelaer, and T.~Stelzer, ``{MadGraph 5
  : Going Beyond},'' {\em JHEP} {\bf 1106} (2011) 128,
  \href{http://www.arXiv.org/abs/1106.0522}{{\tt 1106.0522}}.

\bibitem{Alwall:2007st}
J.~Alwall, P.~Demin, S.~de~Visscher, R.~Frederix, M.~Herquet, {\em et al.},
  ``{MadGraph/MadEvent v4: The New Web Generation},'' {\em JHEP} {\bf 0709}
  (2007) 028, \href{http://www.arXiv.org/abs/0706.2334}{{\tt 0706.2334}}.

\bibitem{Djouadi:2002ze}
A.~Djouadi, J.-L. Kneur, and G.~Moultaka, ``{SuSpect: A Fortran code for the
  supersymmetric and Higgs particle spectrum in the MSSM},'' {\em
  Comput.Phys.Commun.} {\bf 176} (2007) 426--455,
\href{http://www.arXiv.org/abs/hep-ph/0211331}{{\tt hep-ph/0211331}}.
%%CITATION = HEP-PH/0211331;%%.

\bibitem{Nadolsky:2008zw}
P.~M. Nadolsky, H.-L. Lai, Q.-H. Cao, J.~Huston, J.~Pumplin, {\em et al.},
  ``{Implications of CTEQ global analysis for collider observables},'' {\em
  Phys.Rev.} {\bf D78} (2008) 013004,
\href{http://www.arXiv.org/abs/0802.0007}{{\tt 0802.0007}}.
%%CITATION = ARXIV:0802.0007;%%.

\bibitem{Beenakker:1998wi}
{\em {SUSY particle production at the Tevatron}}.
\newblock
1998.
\newblock
%%CITATION = HEP-PH/9810290;%%.

\bibitem{Beenakker:1997ut}
W.~Beenakker, M.~Kr\"amer, T.~Plehn, M.~Spira, and P.~M. Zerwas, ``{Stop
  production at hadron colliders},'' {\em Nucl. Phys.} {\bf B515} (1998) 3--14,
\href{http://www.arXiv.org/abs/hep-ph/9710451}{{\tt hep-ph/9710451}}.
%%CITATION = HEP-PH/9710451;%%.

\bibitem{Huitu:2010ad}
K.~Huitu, S.~Kumar~Rai, K.~Rao, S.~D. Rindani, and P.~Sharma, ``{Probing top
  charged-Higgs production using top polarization at the Large Hadron
  Collider},'' {\em JHEP} {\bf 04} (2011) 026,
\href{http://www.arXiv.org/abs/1012.0527}{{\tt 1012.0527}}.
%%CITATION = 1012.0527;%%.

\bibitem{Rindani:2011pk}
S.~D. Rindani and P.~Sharma, ``{Probing anomalous tbW couplings in single-top
  production using top polarization at the Large Hadron Collider},''
  \href{http://www.arXiv.org/abs/1107.2597}{{\tt 1107.2597}}.

\bibitem{LHCC_stop}
{\bf ATLAS} Collaboration, A.~Hoecker, ``{ATLAS Status and Recent Physics
  Highlights},'' 2012.
\newblock {Talk presented at LHCC meeting, CERN, Dec. 2012.}

\end{thebibliography}\endgroup
\end{document}